\documentclass[prd,twocolumn,amsmath,nofootinbib]{revtex4}
\usepackage{graphicx}

\begin{document}

\def \sub{\ell m \omega}

\title{New Generic Ringdown Frequencies at the Birth of a Kerr Black Hole}
\author{Aaron Zimmerman}
\email{aaronz@caltech.edu}
\author{Yanbei Chen}
\email{yanbei@tapir.caltech.edu}
\affiliation{Theoretical Astrophysics 350-17, California Institute of Technology, Pasadena, California 91125}
\date{\today}

\begin{abstract}

We discuss a new ringdown frequency mode for vacuum perturbations of the Kerr black hole. We evolve initial data for the vacuum radial Teukolsky equation using a near horizon approximation, and find a frequency mode analogous to that found in a recent study of radiation generated by a plunging particle close to the Kerr horizon. We discuss our results in the context of that study. We also explore the utility of this mode by fitting a numerical waveform with a combination of the usual quasinormal modes (QNMs) and the new oscillation frequency.
\end{abstract}

\maketitle

\section{Introduction}  

Black holes are born when a massive star exhausts its nuclear burning processes, leading to a runaway collapse where gravity dominates over all other interactions. They can also be produced by the merger of binary systems containing compact stellar remnants, such as neutron stars or smaller black holes. Stellar collapse and binary mergers resulting in black holes are astrophysical processes where it is expected that gravitational effects are strong, resulting in regions of high curvature. Observations of such processes would provide a strong test of General Relativity.

Gravitational wave astronomy will provide a powerful tool for investigating astrophysical processes involving highly curved regions of spacetime. In the absence of external fields and matter, black hole binary mergers are completely invisible in the electromagnetic spectrum, and no light can reach observers from the interior of a massive star undergoing core collapse. In these situations gravitational waves are expected to carry away as much as a few percent of the total mass energy of the system, and can provide direct information about these otherwise unobservable events (see e.g \cite{WaveReview}).

In this study, we focus on the gravitational wave signal produced in the final stages of the birth of a black hole, when the gravitational waves can be described using linear perturbation theory on the Kerr spacetime. Measurement of such waves could provide a key test of the so called ``No-Hair Theorem'' of General Relativity. The No-Hair Theorem is the statement that stationary black hole spacetimes are described completely using only a few parameters, namely mass, spin, and electric charge. This theorem has been proved for the case of Einstein-Maxwell black hole solutions, through the uniqueness theorem for the Kerr-Newman black holes \cite{Mazur}. Thus, when a black hole is born in a merger or stellar collapse, the resulting object must radiate away all of its multipole moments $\ell \geq 2$ over the course of a ringdown phase. This phase involves emission of gravitational waves in a well known spectrum of exponentially decaying frequency modes, called the quasinormal modes (QNMs) \cite{Nollert1999,Kokkotas,Berti}, and also late time ``tails'' which have a power-law fall-off in time. Observed deviation from QNM oscillation in the ringdown phase would be indicative that the spacetime is not represented by perturbations on a Kerr spacetime, and so would be a violation of the No-Hair Theorem \cite{DreyerTest,BertiLisaTest,BertiFiltering,YunesPPE}.

In addition to this test, detailed study of QNM ringdown is a key component in detection of gravitational waves in the first place. Accurate theoretical and numerical gravitational waveforms are necessary for the success of the method of matched filtering, which will be used to extract the faint signal from the noise in these experiments. Matched filtering uses a gravitational wave template to filter the noise and determine if the wave is present. Accurate modeling of the ringdown phase is then necessary to build useful theoretical templates.

In this study we focus on black hole mergers, which provide a cleaner system with definite numerical predictions, and for which the possibility of detection is higher. The recent strides in numerical relativity  \cite{Pretorious} have allowed several groups to solve the problem of binary inspiral and merger completely for the first time (see \cite{HinderReview,CentrellaReview} for recent reviews). Such simulations have provided enormous insight into binary mergers, and indeed they can serve as a test bed for the theory of black hole perturbations, in addition to providing complete theoretical gravitational wave templates. However, the computational expense of such simulations prohibits their use in generating a large bank of templates for use in matched filtering. As such, a three stage, semianalytic approximation scheme has been developed to treat binary inspirals. This method has the advantage of reducing the computational expense for template generation. Also, analytic methods help to build intuition into the physical processes of the merger.

 The first stage is the long, quasistatic decay of the orbit of the binary, which is treated using the Post Newtonian approximation to General Relativity. The next phase is the rapid merger of the binary, requiring full numerical treatment (though various methods have been employed to approximate the entire merger, e.g. \cite{MergersAppx,NicholsChen}). Finally, once the two compact objects are surrounded by a common horizon, the system can be approximated by the evolution of perturbations of the final Kerr spacetime. The radiation generated in this phase is governed by the Teukolsky equation \cite{Teukolsky}. The QNM frequencies are given by the allowed spectrum of the Teukolsky equation, when physically appropriate boundary conditions are imposed. The QNMs are located at the poles in the Green's function of the radial Teukolsky equation, and are found using a variety of methods (see e.g. \cite{Berti} for a recent review).

Generally, it has been assumed that the QNMs make up the entire spectrum of oscillations during the ringdown phase after merger. Here we seek a new frequency, characterized by the properties of the Kerr horizon. We are inspired by a study by Mino and Brink \cite{MinoBrink}, which investigated the radiation of a point particle falling into a Kerr black hole, using a near horizon expansion to find the radiation analytically.  As the infalling particle approaches the horizon, its trajectory in Boyer-Lindquist coordinates asymptotes to pure angular motion around the black hole with frequency $\Omega_H$,
\begin{equation}
\label{eq1.1}
\Omega_H =\frac{a}{2Mr_+} \ ,\quad r_+ = M + \sqrt{M^2-a^2}\  .
\end{equation}
Here $a$ is the spin parameter, $M$ is the mass, and $r_+$ is the Boyer-Lindquist radius of the outer horizon. From the viewpoint of an observer at infinity, the particle is frozen at the horizon, corotating with it and sourcing radiation at its rotation frequency. Calculations by Mino and Brink show that the radiation arrives at future null infinity with an exponential decay, 
\begin{equation}
\label{eq1.2}
\Psi_4 \sim e^{-im\Omega_H t-g_H t} \ .
\end{equation}
Here $\Psi_4$ encodes the out-going radiation, as discussed fully in Section II, and $m$ is the azimuthal quantum number of the radiation. The decay rate $g_H$ is the surface gravity, given by 
\begin{equation}
\label{eq1.3}
g_H = \frac{ \sqrt{M^2-a^2}}{2Mr_+} .
\end{equation}

The frequency here does not depend on details of the particle's energy and momentum, because the particle's late-plunge trajectory is essentially universal in the Boyer-Lindquist coordinate system. This suggests that this oscillation mode, the ``horizon mode,'' may be more general than this single case considered by Mino and Brink. If we take the naive point of view that the late stages of the merger can be approximated by gravitational perturbations falling onto a final black hole, then we have a situation where the infalling perturbations will source out-going waves like point particles. Though this viewpoint is crude, it does suggest a search for this new frequency mode in post-merger ringdowns.

In this paper we will argue for the existence of a horizon mode (HM) with a frequency of $m \Omega_H$ and a decay constant which we find to be an integer multiple of $ g_H $. We find that the particular decay rate depends on our model for how the spacetime transitions from the nonlinear merger into the regime of first order perturbations on the Kerr spacetime. 

This paper is organized as follows: Section IIA provides a simple argument for the presence of this frequency. In Section IIB we derive this mode through a direct construction, using a simple model for the transition from merger to ringdown. In Section IIC, we explore the consequences of a different model for the transition. In Section III, we reconcile our results with those of Mino and Brink. In order to test the utility of this new HM, in Section IV we use the HM in combination with the QNMs to fit a waveform generated by full numerical general relativity, and compare fits that include the HM to fits with the QNMs alone.

\section{The Near Horizon Approximation}

We first present a heuristic argument for the presence of a HM analogous to that of equation \eqref{eq1.2} in the solutions to the Teukolsky equation. We then derive the HM by evolving initial data for the Teukolsky equation in a near horizon approximation. Finally, we investigate the consequences of a different model for the transition of the spacetime into the regime of linear perturbation theory.

\subsection{Simple Argument for a Horizon Mode}

In the Boyer-Lindquist (BL) coordinate system, the components of the Weyl tensor which represent out-going perturbations of the Kerr spacetime are represented compactly by the Newman-Penrose curvature scalar,
\begin{equation}
\label{eq2.1}
\Psi_4 = C_{\alpha\beta\gamma\delta}n^\alpha \bar{m}^\beta n^\gamma \bar{m}^\delta \ .
\end{equation}

Note that we use a metric signature of $(- + + +)$, and use the appropriate conventions of \cite{FrolovNovikov} for Newman-Penrose (NP) quantities such as $\Psi_4$. We use the Kinnersley null tetrad \cite{Kinnersley}, 
\begin{eqnarray}
\label{eq2.2}
l_{\rm K}^\mu &=&\biggl(\frac{r^2+a^2}{\Delta},1,0,\frac{a}{\Delta}\biggr) \ ,\\
\label{eq2.3}
n_{\rm K}^\mu&=&\frac{1}{2 \Sigma}\biggl(r^2+a^2,-\Delta,0,a \biggr)\ ,\\
\label{eq2.4}
m_{\rm K}^\mu &= & \frac{- \bar\rho}{\sqrt{2}}\biggl(ia \sin\theta,0,1,\frac{i}{\sin\theta}\biggr)
\end{eqnarray}
where
\begin{eqnarray}
\label{eq2.5}
\Delta &=&r^2-2Mr+a^2 \ ,\\
\label{eq2.6}
\rho &=& -\frac{1}{r-ia\cos\theta}\ ,\\
\label{eq2.7}
\Sigma &=& r^2+a^2\cos^2\theta\,,
\end{eqnarray} 
and the overbar represents complex conjugation.

With these choices, $\Psi_4$ satisfies a separable linear wave equation \cite{Teukolsky}, and can be written as
\begin{equation}
\label{eq2.8}
\Psi_4(t,r,\theta, \phi) = \rho^4 \int d\omega \sum_{\ell m} e^{-i \omega t + i m \phi} R_{\sub}(r) S_{\sub} (\theta) \ .
\end{equation}
Here, $S_{\sub} (\theta)$ are the spin-weighted spheroidal harmonics \cite{FackerellCrossman, TeukolskyPress1}, with the appropriate spin weight for $\Psi_4$, $ s = -2$. In the limit $\omega \to 0$, they reduce to the spin-weighted spherical harmonics $ _{-2}Y_{\ell m} (\theta, \phi) $. The radial function $R_{\sub}(r)$ is the solution to the radial Teukolsky equation \cite{Teukolsky}.  Note that $\Psi_4$ vanishes in the background Kerr spacetime, and its perturbed value is independent of tetrad perturbations and gauge transformations.  

We can see the relationship between the scalar $\Psi_4$ and the out-going gravitational waveform via its asymptotic form near future null infinity. In this limit, for asymptotically flat spacetimes (see, e.g. \cite{Schnittman}), 
\begin{equation}
\label{eq2.9}
\Psi_4(r\to \infty) = - \partial_t^2( h_+ - ih_{\times}) \ .
\end{equation}
Here the $+$ and $\times$ indicate the polarization of the gravitational waves.

In the BL coordinates, using the Kinnersley tetrad, the asymptotic behavior near the horizon of the two homogeneous radial solutions are 
\begin{equation}
\label{eq2.10}
R_{\sub}(r) \sim 
\left\{
\begin{array}{cl}
e^{i k r^*} & \mbox{out-going} \\
\\
\Delta^2 e^{-i k r^*} & \mbox{in-going}
\end{array}
\right. \ ,
\end{equation}
with $k = \omega - m \Omega_H$. Together with the separation of $\Psi_4$, equation \eqref{eq2.8}, these solutions are associated with out-going and in-going radiation at the horizon. The tortoise coordinate $r^*$ is defined by $dr^*/dr = (r^2+a^2)/\Delta$. Note that $r^* \to - \infty$ as $ r \to r_+$.

One of these two solutions is selected out as unphysical, based on its behavior near the event horizon. Here we repeat an argument first presented by Teukolsky \cite{Teukolsky}. We demand that fields neither vanish exactly nor diverge at the horizon when measured by a physical observer. Near the horizon, the trajectory of any freely falling observer will approximately match that of an in-going null geodesic (see e.g. \cite{MinoBrink,MTW}), independent of the observer's energy or angular momentum, with
\begin{equation}
\label{eq2.15}
t + r^*= {\rm const}, \; \theta={\rm const}, \; \phi=\Omega_H t\,.
\end{equation}
This means that for distant observers, the infalling trajectory does not appear to enter the horizon, but instead asymptotes to it while circulating around the horizon with a constant frequency. A more natural set of coordinates is in-going Kerr coordinates, $(v,r,\theta, \tilde{\phi} )$, where $ dv = dt + dr^*$, and $d\tilde{\phi} = d\phi + a \  dr^*/(r^2+a^2)$. In-going null geodesics lie on lines of constant $v$ and $\tilde{\phi}$. 

In this in-going coordinate system, the metric does not become singular at the horizon. However, even in these coordinates, the Kinnersley tetrad used to define $\Psi_4$ becomes singular at the horizon. This can be repaired by using a Lorentz transform to boost into the reference frame of an infalling observer who carries a nonsingular tetrad, namely
\begin{equation}
\label{eq2.11}
\mathbf{l}_{\rm in} = \frac{\Delta}{2(r^2+a^2)} \mathbf{l}_{\rm K}\,,\quad \mathbf{n}_{\rm in} = \frac{2(r^2+a^2)}{\Delta} \mathbf{n}_{\rm K}
\end{equation}
Here the subscript ``in'' indicates the regular in-going tetrad. The tetrad, written in the in-going Kerr components, is
\begin{eqnarray}
\label{eq2.12}
l_{\rm in}^{\mu}&  = & \biggl( 1, \frac{\Delta}{ 2 (r^2+a^2)}, 0 , \frac{a}{r^2+a^2} \biggr)\  ,
\nonumber \\
 n_{\rm in}^{\mu} & = & \biggl( 0, - \frac{r^2 +a^2}{\Sigma},0,0 \biggr) \ , 
\nonumber \\
m_{\rm in}^{\mu} & = & \frac{-\bar{\rho}}{\sqrt{2}}\biggl(  i a \sin \theta, 0 , 1, \frac{i}{\sin \theta} \biggr) \ .
\end{eqnarray}
With this tetrad, the physical observer measures a curvature scalar of 
\begin{equation}
\label{eq2.16}
\Psi_4^{\rm in} = \left[ \frac{2(r^2+a^2)}{\Delta} \right]^2 \Psi_4 \ .
\end{equation}
As a consequence, the two radial solutions in equation \eqref{eq2.10} correspond to  $\Psi_4^{\rm in} \sim e^{-i\omega v}e^{im\tilde{\phi}}$ and $\Psi_4^{\rm in} \sim \Delta^{-2} e^{-i\omega v}e^{im\tilde{\phi}}e^{2ikr^*}$. The second diverges at the horizon, and so is selected as unphysical. In other words, if waves emerge from the horizon, then the in-going observers will see a diverging curvature due to blueshift effects. Note that while it is a particular observer that carries the tetrad of equation \eqref{eq2.12} near the horizon, these general results hold for all physical observers. This is because the tetrad that another physical observer carries can be related to that in equation \eqref{eq2.12} through nonsingular Lorentz transformations.

However, if the frequency $\omega$ is complex, this divergence can be removed. To see this, we let $\omega = m \Omega_H - i \gamma$, and seek an appropriate $\gamma$ (the real part of the frequency is chosen so that the observer does not measure increasingly rapid oscillations when approaching the horizon). We note that near the horizon, $\Delta \sim e^{2 g_H r^*} $.  For the physical observer, then,

\begin{equation}
\label{eq2.17}
\Psi^{\rm in}_4  \sim  \exp \biggl( -i \omega v + i m \tilde {\phi} -4 g_H r^*  + 2 \gamma r^*  \biggr) \ .
\end{equation}

We see from this that if 
\begin{equation}
\label{CriticalDecay}
\gamma = 2 g_H \,, 
\end{equation}
this solution remains regular at the horizon, and the solution decays along the observer's worldline in just such a way that the growth is compensated for. This particular frequency is selected out by a physically allowed solution of the Teukolsky equation, and so we must consider its place in the usual QNM spectrum. Modes with $\gamma > 2 g_H$ decay exponentially for the physical observer. In this sense, the decay rate of equation \eqref{CriticalDecay} gives the least damped, physically reasonable mode in this simple argument.

\subsection{The Near Horizon Limit: The $\Sigma$ Boundary Model}

With our heuristic argument in hand, we now derive the HM using a specific ansatz. In order to keep our results relevant to the problem of compact binary inspiral, we consider the following model. We imagine that to the future of some spacelike hypersurface $\Sigma$, the spacetime can be described with linear perturbation theory on the Kerr spacetime, while to the past of $\Sigma$ the spacetime may be nonlinear. We denote $\Sigma$ by setting the Boyer-Lindquist time coordinate $t=0$. The past of $\Sigma$ represents the inspiral and merger phases of binary coalescence. To the future of $\Sigma$ we can use the Teukolsky equation to evolve initial perturbations on $\Sigma$ forward. A similar ansatz has been used in the Close Limit approach \cite{CloseLimit}, and Lazarus project \cite{Lazarus} which used numerical integration of the Teukolsky equation to evolve initial data on an initial time slice \cite{CloseLimitInt, Lazarus}. We refer to our model as the $\Sigma$ boundary model.

Given this ansatz, we evolve initial data $\psi |_{t=0} \equiv \rho^{-4}\Psi_4|_{t=0}$ and $\partial_t \psi|_{t=0}$, using the Green's function. The full details of the analysis are presented in Appendix A. A near horizon expansion allows us to obtain part of the evolution analytically. Physically, we postulate that just after the merger of the binary, the perturbations are concentrated in a small region near the horizon, so that the initial data used in the Green's function evolution only has support in a small region near the horizon. This expansion allows us to use the asymptotic form for the Green's function, and to keep terms only to first order in $\epsilon \equiv (r-r_+)/r_+ \ll 1 $. While we focus on the physical picture where the perturbations are concentrated near the horizon, our results hold for the evolution of the initial data which is near the horizon even if the data on the initial surface extends to large $r$. In addition, due to redshift effects, this finite region near the horizon produces (decaying) radiation over an infinite region of null infinity.

Specifically, let $\psi|_{t=0}$ and $\partial_t \psi|_{t=0}$ be nonzero only between $r_+$ and $(1 + \xi)r_+$, with $\xi \ll 1$.  We truncate the integrals of the Green's function over the initial perturbation to this small region. To first order in distance from the horizon, $\Delta \approx 2Mr_+\kappa\epsilon$, with $\kappa \equiv \sqrt{1 - a^2/M^2}$. From Appendix A, equation (\ref{eqa.23}), we have that $\Psi_4$ takes the form of equation (\ref{eq2.8}) with the radial function $R_{\sub}(r)$ given to leading order in $\epsilon$ by equation (\ref{eqa.24}),
\begin{eqnarray}
\label{eq2.18}
R_{\sub} (r) & \approx & - \int_{r_+}^{(1+\xi)r_+} dr' \biggl[ \frac{\beta_{\sub}(r') + i \omega \alpha_{\sub}(r') }{2 M r_+ (\kappa \epsilon)^3} 
\nonumber \\ && \qquad 
+ \frac{(2 M \kappa + i m a) \alpha_{\sub}(r')}{2(M r_+)^2 (\kappa \epsilon)^3} \biggr] \tilde{G}_{\sub}(r,r') \ ,
\nonumber \\
\end{eqnarray}
where the functions $\alpha_{\sub}(r)$ and $\beta_{\sub}(r)$ can be found from the initial data, using equations (\ref{eqa.15}), (\ref{eqa.16}), (\ref{eqa.21}), and (\ref{eqa.22}). The function $\tilde{G}_{\sub}(r, r') $ is the frequency domain radial Green's function.

 In order to evaluate this expression, we insert the explicit form of the radial Green's function,
\begin{eqnarray}
\label{eq2.19}
\tilde{G}_{\sub}(r, r') & = & 
\frac{1}{W_{\sub}} \Biggl\{ 
{R^{\rm up}_{\sub}(r) \ R^{\rm in}_{\sub}(r') \indent r' < r
\atop R^{\rm in}_{\sub}(r) \ R^{\rm up}_{\sub}(r') \indent r' > r} \ .
\nonumber \\
\end{eqnarray}
The functions $R^{\rm up}_{\sub}$ and $ R^{\rm in}_{\sub}$ are two homogeneous solutions to the radial Teukolsky equation, with the up-mode (no radiation from past null infinity) and in-mode (no radiation from the past horizon) boundary conditions, respectively. They have the asymptotic forms
\begin{eqnarray}
\label {eq2.20}
R^{\rm in}_{\sub}(r) & \to &
 \Biggl\{  
{B^{\rm trans}_{\sub} \Delta^2 e^{-i k r^*} 
\quad
r \to r_+
\atop 
B^{\rm ref}_{\sub} e^{i \omega r^*} + B^{\rm in}_{\sub} r^{-1} e^{-i \omega r^*} \quad 
r \to \infty} \,,
\nonumber \\ 
\\
\label{eq2.21}
R^{\rm up}_{\sub}(r) &\to&
 \Biggl\{  
{C^{\rm up}_{\sub} e^{ikr^*} + C^{\rm ref}_{\sub} \Delta^2 e^{-ikr^*} \quad r \to r_+
\atop 
C^{\rm trans}_{\sub} r^3 e^{i \omega r^*} 
\qquad 
r \to \infty} \,,
\nonumber
\\ 
\end{eqnarray}
where the Wronskian $W_{\sub}$ is given by $ W_{\sub} = 2 i \omega B^{\rm in}_{\sub} C^{\rm trans}_{\sub}$.

Since we are interested in the waves at infinity, $r \to \infty$, we insert the appropriate asymptotic expression for $R^{\rm in}_{\sub}$. Also, since the integral extends only over the near horizon region, we insert the asymptotic expression of $R^{\rm in}_{\sub} \to B^{\rm trans}_{\sub} \Delta^2 e^{-ikr^*} \approx B^{\rm trans}_{\sub} (2 M r_+ \kappa \epsilon)^2 e^{-ikr^*}$. Thus,
\begin{equation}
\label{eq2.22}
R_{\sub}(r\to \infty) \approx \frac{B^{\rm trans}_{\sub}}{2 i \omega B^{\rm in}_{\sub}} r^3 e^{i \omega r^*} Z_{\sub} \ ,
\end{equation}
\begin{eqnarray}
\label{eq2.23}
Z_{\sub} &= & - \frac{2}{\kappa}
\int_{r_+}^{(1+\xi)r_+} dr \ \epsilon^{-1} e^{-i k r^*}
\biggl[Mr_+ \beta_{\sub}(r) 
\nonumber \\ & & \qquad
+ (i M  \omega r_+ +2 M \kappa + i m a ) \alpha_{\sub}(r)\biggl] \ . \nonumber \\
\end{eqnarray}
In order to complete the integration, we need to know how the initial data behaves to leading order in $\epsilon$. We can write to leading order $\alpha_{\sub}(r) \approx \alpha^0_{\sub} \epsilon^n$ and $\beta_{\sub}(r) \approx \beta^0_{\sub} \epsilon^p$. We find the leading order powers $n$ and $p$ by essentially repeating the argument given in Section IIA, with some additional care. 

Here we cannot follow a single observer who falls past the horizon, since we wish to know the behavior of $\Psi_4$ on the initial slice $\Sigma$. We consider instead a family of accelerated observers who cross the surface $t=0$ at a variety of radii, extending all the way to the horizon but not penetrating it. We choose the four-velocities of these observers at this initial surface to vary with $r$, $\mathbf{u}(r)$, such that all members of this family carry the regular tetrad \eqref{eq2.12}. These observers measure an initial perturbation in their own frames, and for an observer at some $r$ the measured perturbation is related to that expressed in the Kinnersley tetrad by $\Psi^{\rm in}_4 = [2(r^2+a^2)/\Delta]^2 \Psi_4 \sim \epsilon^{-2} \Psi_4$. Since all the observers are physical, we expect that the perturbation they measure does not diverge as we take the limit $r \to r_+$, moving along the family of observers. However, this requires that $\Psi_4 \sim \epsilon^2$. Thus, we require $n=2$, $p=2$ on this interval.

Inserting this into \eqref{eq2.23} we have 
\begin{eqnarray} 
\label{eq2.24}
Z_{\sub} &=&  -\biggl[Mr_+ \beta^0_{\sub} 
+ (i M \omega r_+ +2 M \kappa + i m a ) \alpha^0_{\sub}\biggl] 
\nonumber \\ & & \times
\frac{2}{\kappa} \int_{r_+}^{r_+(1+\xi)} dr  \ \epsilon \ e^{-ikr^*} \nonumber \\
&=& \biggl[Mr_+ \beta^0_{\sub} 
+ (i M \omega r_+ +2 M \kappa + i m a ) \alpha^0_{\sub}\biggl] 
\nonumber \\ & & \times
\frac{2}{\kappa}
 \int_{r_+}^{r_+(1+\xi)} dr \ \epsilon^{1-i k r_+/\kappa} \ .
\end{eqnarray}
Focusing on the integral, which we denote $I$, we have
\begin{equation}
\label{eq2.25}
I = \frac{i \kappa}{\omega- (m \Omega_H - 4 i g_H)} e^{(2-ik/2g_H) \ln \xi} \ .
\end{equation}
Here, there is a pole in the denominator which will select out the frequency
\begin{equation}
\label{sigmafreq}
\omega_H = m \Omega_H - 4 i g_H \,, \qquad (\Sigma \ {\rm Model})
\end{equation}
 when equation \eqref{eq2.8} is integrated over $\omega$. 

Combining equations \eqref{eq2.8}, \eqref{eq2.22}, \eqref{eq2.24}, and \eqref{eq2.25}, we integrate over $\omega$. We close the contour in the lower half plane, selecting out the poles of equation \eqref{eq2.22} by the residue theorem. The zeros of $B^{\rm in}_{\sub}$ comprise one set of poles in the lower half plane, and these poles give the usual QNM frequencies. We wish to focus on the contribution from the additional pole at $\omega_H$, and so from here we ignore the poles coming from $B^{ \rm in}_{\sub}$. Similarly, we will not consider here the influence of the pole at $\omega = 0$ (actually part of a branch cut along the negative imaginary axis, which generates the late-time power-law tails). Also, we will  not consider the possible poles in the terms $\alpha^0_{\sub}$ and $\beta^0_{\sub}$, which receive their frequency dependence from the projection of the initial data onto the spheroidal harmonics. 

 This integral converges for $t-r^*+ \ln \xi / 2 g_H > 0$; otherwise, we must close the contour in the upper half plane and the integral vanishes. Noting that as $r \to \infty$, $\rho \to - r^{-1}$, we have
\begin{eqnarray}
\label{eq2.26}
\Psi_4 & = & \frac{1}{r} \sum_{\ell m} \tilde{Z}_{\sub} e^{-i \omega_H (t- r^*)+ i m \phi} \ S_{\sub}(\theta) \  
\nonumber \\ & & \times
H\biggl(t-r^* + \frac{\ln \xi} {2 g_H} \biggr) \ ,
\\
\label{eq2.27}
\tilde{Z}_{\sub} & = & \frac{2 \pi B^{\rm trans}_{\sub}}{\omega_H B^{\rm in}_{\sub}} \biggr[ (M \omega_H r_+ + ma - 2 i M \kappa) \alpha^0_{\sub}
\nonumber \\ & & \qquad \qquad \qquad
 - i M r_+ \beta^0_{\sub}\biggr]  \ ,
\end{eqnarray}
where $H$ is the unit step function. In the above, all frequencies are to be evaluated at $\omega_H$ from \eqref{sigmafreq}. As $\xi \ll 1$, the waves at infinity appear at late retarded times. This sharp turn-on of the wave is an artifact of our truncation of the integral at $(1 + \xi) r_+$. A smoother falloff of the initial data with increasing radii would result in a smoother turn on of the wave at infinity. These waves at early times are sourced by initial data on $\Sigma$ which cannot be evolved using the near horizon approximation. We see also that the waves continue to reach infinity for all retarded times after the turn on. As mentioned previously, this is due to redshift effects near the horizon, which stretch the radiation from the finite near horizon region out over an infinite region of null infinity.

\subsection{The Characteristic Boundary Model}
\label{CharModel}

While the frequency of the radiation in equation \eqref{eq2.25} matches the result of our heuristic argument, the decay rate does not. The decay rate is determined by the radial behavior of the initial data, as we can see in equations \eqref{eq2.24} and \eqref{eq2.25}. In order that $ \omega_H = m \Omega_H - 2 i g_H $, the initial data would need to behave as $\Psi_4|_{t=0} \sim \epsilon$, which we have argued against based on our requirement that physical observers near the horizon measure nonsingular initial data. However, a change of our ansatz shows that the initial data can be proportional to $\epsilon$ and still represent physical perturbations. In this case the surface bounding the regime of linear perturbation theory is an in-going null surface, instead of a surface of constant $t$. This differs from the $\Sigma$ boundary model, and so differs from the Close Limit Approach. We will refer to this second model as the characteristic boundary model.

\begin{figure}[t] 
\begin{center}$
\begin{array}{c}
  \includegraphics[width=3.375in, keepaspectratio]{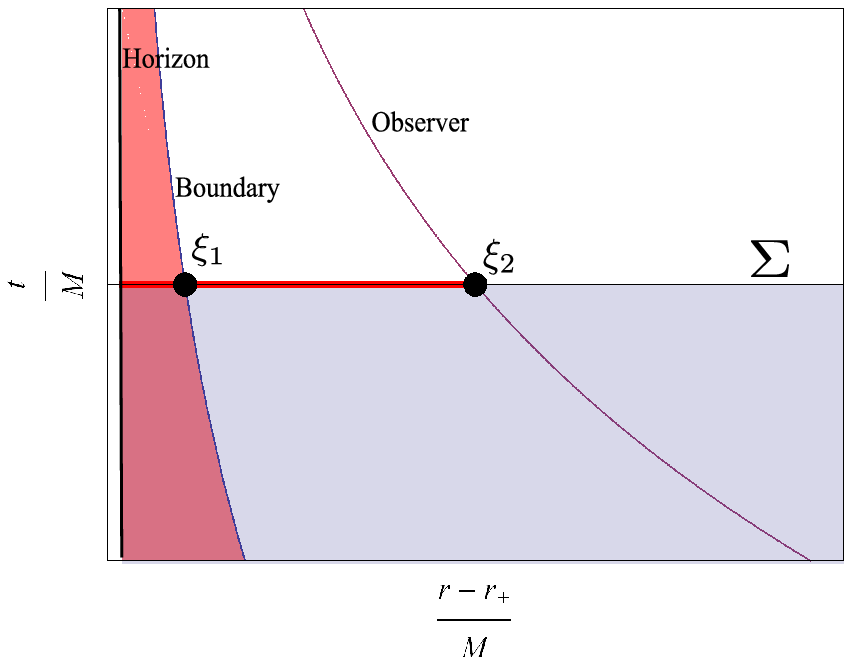} \\
  \includegraphics[width=3.375in, keepaspectratio]{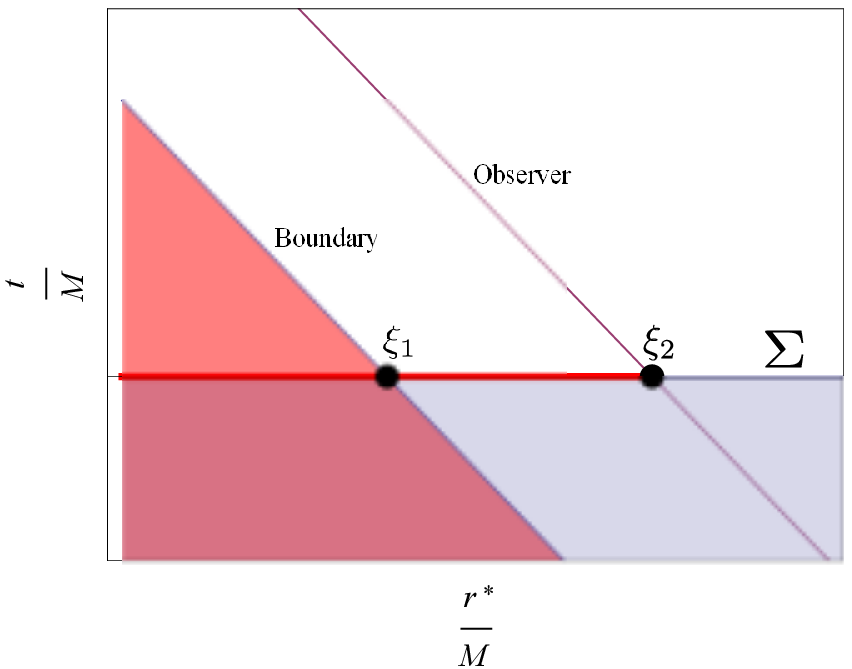}
\end{array}$
\end{center} 
  \caption{(color online) Comparison of the $\Sigma$ and characteristic models. In both figures, the trajectory of the observer discussed and the characteristic boundary surface are plotted, with $(\theta, \phi)$ suppressed. The thick line is the interval where the initial data is nonzero, and the two points are $\xi_1$ and $\xi_2$, respectively. The shaded regions correspond to the nonlinear regime of each model.  Top: Comparison in BL coordinates of the two models, with the event horizon illustrated. Bottom: Comparison in $(t,r^*)$ coordinates, where the horizon is pushed to $r^* \to \infty$.}
  \label{fig:Models}
\end{figure}

As the spacetime transitions into the linear regime, the nonlinear perturbations radiate away towards infinity or down into the black hole. We imagine that the regions of nonlinear evolution are bounded by characteristics of the linear wave equation. This is a more physically motivated assumption than a transition in spacetime properties along the entire surface $\Sigma$. The in-going characteristics have a trajectory $r(t) = r_+(1+e^{-2 g_H (t-t_0)})$ \cite{MinoBrink}, where $t-t_0 \gg 1$. The comparison of these two models in both BL $(t,r)$ coordinates and the tortoise $(t, r^*)$ coordinates is given in Figure \ref{fig:Models}. In the $(t, r^*)$ coordinates, we see that the horizon is pushed to $r^* \to - \infty$., and that the initial data of the $\Sigma$ model is stretched out onto an infinite interval in $r^*$. We see in both figures that the horizon is hidden behind the boundary characteristics, and our previous argument for the radial dependence of the initial data on $\Sigma$ no longer holds. We must find a new way to determine the $r$ dependence of initial data in this model.

We will again evolve initial data on the constant time slice $\Sigma$, this time with support only on a small interval outside the boundary characteristics. At time $t=0$, we set the inner boundary of the initial data set to be at $r_+(1+\xi_1)$, and the outer boundary at $r_+(1+\xi_2)$. We will use the same physical observers as before. To first order their trajectories are lines of constant $v$ and $\tilde{\phi}$, just like the in-going boundary characteristics. The physical observer who passes $r_+(1+\xi_2)$ at $t=0$ has a trajectory $r_{\rm obs}(t) = r_+(1+e^{-2 g_H (t-t_{\rm obs})})$. At $t=0$ the observer measures an initial perturbation $\Psi^{\rm in}_4 \sim \xi_2^{-2} \Psi_4(r_{\rm obs}(0))$. At a later time $t$, the observer measures $\Psi^{\rm in}_4 \sim \xi_2^{-2} e^{4 g_H t} \Psi_4(r_{\rm obs}(t))$, and so the measured perturbation grows exponentially in time. However, the perturbation will also decay in time due to its evolution. We insist then that the decay be such that this observer (and similarly, all of the physical observers near the horizon) do not experience an exponentially growing perturbation. This will set the behavior of $\Psi_4$ on the initial surface.

We examine then the perturbation as measured by the observer along his trajectory. Given data that behaves as $\alpha_{\sub}= \alpha^0_{\sub} \epsilon^n $ and $\beta_{\sub}= \beta^0_{\sub} \epsilon^n $ on the initial surface, we again combine equations \eqref{eq2.18}- \eqref{eq2.21}, this time taking the asymptotic limits as $r' \to r_+$ and $r \to r_+$. We focus on the out-going solution only, since the observer will not measure the in-going waves. We find that
\begin{eqnarray}
\label{eq2.28}
R_{\sub}(r) & \sim & \frac{e^{ikr^*}}{\omega - (m\Omega_H - 2 n i g_H)}
\nonumber \\ & & \times
 \biggl( e^{-2g_H(t-t_{\rm obs})(n- ikr_+/\kappa)}-e^{\ln\xi_1(n-ikr_+/\kappa)}\biggr) \ .
\nonumber \\
\end{eqnarray}
Integrating equation \eqref{eq2.8} with this radial function, and noting again that for the observer $v_{\rm obs } = t+r^*$ is constant, so that $t = v_{\rm obs } - r^*$, we have
\begin{equation}
\label{eq2.29}
\Psi_{4}(r_{\rm obs}) \sim e^{-i m \Omega_H t - 4 n g_H t} \biggl[ H(t + t_{\rm obs}) - H \biggl(2t + \frac{\ln \xi_1}{2g_H} \biggr) \biggr] \ . 
\end{equation}
For $n=1$, the decay of the perturbation along the worldline of the observer is just enough to cancel the exponential growth. The initial data on $\Sigma$ can be taken to be proportional to $\epsilon$. 

With the initial data, we can return our attention to the perturbations measured at infinity. Repeating the analysis of Section IIB with this initial data, we have for $r \to \infty$
\begin{eqnarray}
\label{eq2.30}
\Psi_4 & = & \frac{1}{r} \sum_{\ell m} \tilde{Z}_{\sub} e^{-i \omega_H (t- r^*)+ i m \phi} S_{\sub}(\theta)
\nonumber \\ & & \times
\biggl[ H \biggl( t-r^* + \frac{\ln \xi_2} {2 g_H} \biggr)-H \biggl(t-r^* + \frac{\ln \xi_1} {2 g_H} \biggr) \biggr] \ ,
\nonumber \\
\end{eqnarray}
with $\tilde{Z}_{\sub}$ as in equation \eqref{eq2.27} but with $\omega_H$ given by
\begin{equation}
\label{eq2.31}
\omega_H =  m \Omega_H - 2 i g_H \,.  \qquad ({\rm Boundary \, Model})
\end{equation}
 We have recovered the decay rate indicated by the heuristic argument of Section IIA. The difference of step functions here and in equation \eqref{eq2.29} is again due to the sharp truncation of the integral at each end of the interval on $\Sigma$. Now the radiation turns off due to the truncation of the initial data at the boundary characteristic $\xi_1$. At this retarded time the gravitational radiation would give way to radiation sourced by the perturbations in the nonlinear region of spacetime on $\Sigma$. For this second model it seems that a method for evolving data along characteristics would be better suited than evolution from a constant time slice. Such a characteristic evolution has been presented for the Schwarzschild black hole, for example in \cite{Price}. Others \cite{Nullevolution} have presented numerical evolution of characteristics, again for the Schwarzschild black hole. Another possible formulation which would be natural in this context would be the use of an asymptotically hyperboloidal spacelike surface in place of $\Sigma$, as discussed in \cite{HypSlicing}.

\section{Reconciliation with the Mino-Brink mode}
\label{MBCorrect}

In Section IIB and IIC, we saw that the condition that physical observers measure regular curvature near the horizon determines the decay rate for the gravitational radiation at infinity. In fact, the mode of Mino and Brink, with frequency $\omega = m \Omega_H - i g_H$, has a decay rate which violates the regularity conditions discussed in both sections. Its decay rate is too small, for example compared to equations \eqref{sigmafreq} and \eqref{eq2.31} with decay rates $\gamma \geq 2 g_H$. Thus, although the mode found in \cite{MinoBrink} motivates our study, the two results are in disagreement. In this section, we first provide a simple alternative estimate for the expected decay rate of radiation from a point particle, using the notation of Newman and Penrose for convenience \cite {NewmanPenrose}.  With this estimate as motivation, we then find that correction of an error in \cite{MinoBrink} unexpectedly leads to the vanishing of the first order mode discussed there. We conclude that the actual leading order radiation from an infalling point particle has a decay rate which matches our characteristic boundary model in Section IIC. 

\subsection{Point Particle Radiation in Newman-Penrose Formalism}

We wish to calculate the radiation generated by a point particle at the last stages of its plunge into a Kerr black hole. Once again, the radiation is described by the Teukolsky formalism, this time with the appropriate source term for the matter content. However, it will turn out to be convenient to make the near horizon expansion in the Newman-Penrose formalism, in order to obtain an estimate for the behavior of the radiation. As we show in Appendix B, we can write to leading order in $\Delta$ $(\epsilon)$ near the horizon, using the Kinnersley tetrad,
\begin{equation}
\label{eq3.1}
(\hat D \hat \Delta + 4\gamma \hat D )\Psi_4 \approx 4\pi T_4 \ ,
\end{equation}
where
\begin{eqnarray}
\label{eq3.2}
\hat D & = & l_{\rm K}^{\mu} \nabla_{\mu} \ ,
\\
\label{eq3.3}
\hat{\Delta} & = & n_{\rm K}^{\mu} \nabla_{\mu} \ ,
\\
\label{eq3.4}
\gamma & = & \frac{r-M}{2 \Sigma} \ ,
\end{eqnarray}
and where the source term $T_4$ is given by \eqref{eqb.2}.

In addition, we can approximate
\begin{eqnarray}
\label{eq3.5}
\hat D &\approx& \frac{r_+^2+a^2}{2\Delta}(\partial_t +\partial_{r^*}+\Omega_H \partial_\phi) \equiv \frac{r_+^2+a^2}{2\Delta} L_+ ,  \\
\label{eq3.6} 
\hat \Delta &\approx & \frac{r_+^2+a^2}{2\Sigma} (\partial_t -\partial_{r^*}+\Omega_H \partial_\phi)\equiv \frac{r_+^2+a^2}{2\Sigma} L_-  ,
 \end{eqnarray}
and write
\begin{equation}
\label{eq3.7}
M^2 r_+^2 (L_+ L_- +4 g_H L_+ )\Psi_4 \approx 4 \pi \Sigma\Delta T_4 \ .
\end{equation}
In absence of $T_4$, this directly gives the asymptotic in-going and out-going waves of equation \eqref{eq2.10}. Let us specialize $\Psi_4$ to a particular azimuthal quantum number, $m$, and we have
\begin{equation}
\label{eq3.8}
L_{\pm} = \partial_t \pm \partial_{r_*} + i  m \Omega_H \ .
\end{equation}

We turn now to $T_4$. We expand the stress-energy tensor of a point particle in terms of the azimuthal quantum number $m$ to match the expansion of $\Psi_4$ implicit in \eqref{eq3.8}, which gives
\begin{eqnarray}
\label{eq3.9}
T^{\mu\nu} &=&\mu \frac{u^\mu u^\nu}{\sqrt{-g}}\frac{1}{u^t} \delta(r-r(t))\delta(\theta-\theta(t)) e^{im(\phi-\phi(t))} \\
\label{eq3.10}
&=&\mu \frac{u^\mu u^\nu  }{u^t\Sigma} \frac{dr^*}{dr}\delta(r^*-r^*(t))\delta(\chi-\chi(t))e^{im(\phi-\phi(t))} \ , \nonumber \\
\end{eqnarray}
Here we have defined $\chi\equiv \cos \theta$, and used the properties of the delta function. The mass of the particle is given by $\mu$.  If we define $v=t +r^*$ and $u=t-r^*$,  and once again use the properties of the delta function, we can write
\begin{eqnarray}
\label{eq3.11}
\delta(r^* - r^*(t)) & = &\delta\biggl( \frac{u-v}{2}- r^*\biggl(\frac{u-v}{2}\biggr) \biggl)
\\
\label{eq3.12}
& = & 2 \frac{\delta( v- v_0(t))}{du/dt} \ ,
\end{eqnarray}
where $v_0(t)$ is the value at which the argument of the delta function in \eqref{eq3.11} vanishes, which is at first order simply the value $v_0= \mbox{const}$ to which the trajectory asymptotes. Also, to leading order the trajectory will have $ t = -r^*$, so $du/dt = 2$ to leading order.

We have then that
\begin{equation}
\label{eq3.13}
T^{\mu\nu} = \mu \frac{2 u^\mu u^\nu}{u^t\Sigma}\frac{dr^*/dr}{du/dt}\delta(v-v_0(t))\delta(\chi-\chi(t)) e^{im(\phi- \Omega_H t)} \ .
\end{equation}
We must now project $T^{\mu \nu}$ onto the null Kinnersley basis in order to find $T_4$, see \eqref{eqb.2}. This will result in a projection of the four velocity onto the Kinnersley basis, for example with $u_n = u_{\mu} n^{\mu}$,  and in this basis some of the components are vanishing or divergent as $r \to r_+$ (as seen in Section IIA). In order to examine the leading order behavior near the horizon, it is then best to express the four velocity components in terms of the regular, in-going basis, related to the Kinnersley tetrad by \eqref{eq2.11}. In the in-going basis, we have 
\begin{equation}
\label{eq2.13}
 \mathbf{u} = u^l_{\rm in} \mathbf{l}_{\rm in}+ u^n_{\rm in} \mathbf{n}_{\rm in}
+u^m_{\rm in} \mathbf{m}_{\rm in}+u^{\bar m}_{\rm in} \bar{\mathbf{m}}_{\rm in}\,,
 \end{equation}
with $u_{\rm in}^{l,n,m,\bar m}$ all smooth and finite throughout the trajectory --- including on and within the future horizon. The four velocity in the Kinnersley basis can be expressed as
\begin{eqnarray}
\label{eq2.14}
 \mathbf{u}& =& 
 u^l \mathbf{l}_{\rm K}+ u^n \mathbf{n}_{\rm K}
+u^m \mathbf{m}_{\rm K}+u^{\bar m} \bar{\mathbf{m}}_{\rm K}
\\
& = & \frac{\Delta u^l_{\rm in}}{2(r^2+a^2)}   {\mathbf{l}_{\rm K}}+\frac{ 2(r^2+a^2)u^n_{\rm in}}{\Delta} \mathbf{n}_{\rm K}
+u^m_{\rm in} \mathbf{m}_{\rm K}
\nonumber \\ & &
+u^{\bar m}_{\rm in} \bar{\mathbf{m}}_{\rm K} \,.
 \end{eqnarray} 
Finally, we lower the tetrad indices on the components of the four velocity using the null metric
\begin{equation}
\label{eq3.17}
\hat \eta_{ab} = \left(  \begin{array} {cccc}
0 & -1 & 0 & 0 \\
-1 & 0 & 0 & 0 \\
0& 0 & 0 & 1 \\
0& 0 & 1 & 0 \end{array} \right) \ .
\end{equation}
This gives us
\begin{eqnarray}
\label{eq3.14}
u_n u_n &=&\Delta^2 (u^l_{\rm in})^2 \ , \\ 
\label{eq3.15}
u_n u_{\bar m} &=& - \Delta u^l_{\rm in}u^{m}_{\rm in} \ , \\
\label{eq3.16}
u_{\bar m} u_{\bar m}& =& (u^{m}_{\rm in})^2 \ .
\end{eqnarray}

Let us first consider $T_{\bar m\bar m}$, which gives the dominant contribution $T_{4 \bar m \bar m}$ to $T_4$,
\begin{eqnarray}
\label{eq3.18}
T_{4\bar m \bar m} &\approx& - (\hat\Delta+2\gamma)\hat\Delta T_{\bar m\bar m}
\end{eqnarray}
It turns out that to first order $\hat{\Delta} T_{\bar m \bar m} = 0$. We can see this by noting that at first order 
\begin{eqnarray}
\label{eq3.19.1}
\Sigma u^t & \approx & \frac{(r^2+a^2)^2}{\Delta} (E - \Omega_H L_z)
\\ 
\label{eq3.19}
T_{\bar m \bar m} & \approx& \mu \frac{ (u_{\rm in}^m)^2 e^{im(\phi- \Omega_H t)}}{2 M r_+(E- \Omega_H L_z)}\delta(v - v_0) \delta(\chi - \chi_0) \ ,
\end{eqnarray}
with $\chi_0 = \cos \theta_0$ the value of $\chi$ to which the particle asymptotes to at the horizon. Acting on this with $\hat \Delta$ from \eqref{eq3.6}, we get
\begin{equation}
\label{eq3.20}
\hat{\Delta} T_{\bar m \bar m} \sim \frac{r_+^2 + a^2}{2 \Sigma}(-i m \Omega_H T_{\bar m \bar m} + i m \Omega_H T_{\bar m \bar m}) = 0 \ .
\end{equation}
So we see that
\begin{equation}
\label{eq3.21}
\hat \Delta T_{\bar m \bar m } \sim \Delta.
\end{equation}
Physically, this is because $\hat{\Delta}$ takes its derivative almost along the direction of motion, along which $T_{\bar m \bar m}$ does not change to first order.  This  means that the exact contribution of $T_{\bar m \bar m}$ to $T_4$ must be reexamined with other terms included, and its contribution is in fact at the order of $T_{\bar m n}$.  Thus, we expect
\begin{equation}
\label{eq3.20}
(L_+ L_- +4 g_H L_+ )\Psi_4 \sim \Delta ^2 \delta(v-v_0)
\end{equation}

We have a simple scenario: if we integrate across the $v=v_0$ surface, removing the derivatives $\partial_v = (\partial_t + \partial_{r^*})/2$ from the left-hand side, we will have a $u$-dependent $\Psi_4$, which obeys
\begin{equation}
\label{eq3.21}
\Psi_4 \sim \Delta^2 \sim e^{4 g_H r^*} \sim e^{-4g_H t} \ .
\end{equation}
Here we have recovered the decay rate near the horizon discussed in Section IIC. This indicates that the corresponding decay rate of the waves as $r \to \infty$ is that of equations \eqref{eq2.30}-\eqref{eq2.31}. However, this argument lacks the detailed calculations of Mino and Brink, who found a mode with a slower decay rate. Under examination, however, it is an error in \cite{MinoBrink} which leads to a mode with a spuriously low decay rate. We discuss this in the next section.

\subsection{Eliminating the Leading Order Frequency Mode of Mino and Brink}

We turn now to the study by Mino and Brink, which we abbreviate as MB. In this study, the source term is evolved using the Green's function method much as Section II of our study, resulting in an integrand for the integral \eqref{eq2.19} which has a pole at $\omega = m \Omega_H - i g_H$, i.e. equation (MB 3.7). The pole then selects out this oscillation frequency for the out-going radiation at infinity. However, it turns out that this pole is canceled out by terms in the amplitude $\tilde{Z}_{\sub}$, when an error in MB is corrected for. We find this error in going from (MB A14) to (MB B4). The first equation gives a piece of the Fourier decomposition of the source term $T_4$, and is drawn from \cite{MinoPert}, equations (2.21) and (2.25) therein. 

Taking the leading order contribution of (MB A14), we find that (MB B4) should read at leading order
\begin{eqnarray}
\label{eq3.22}
A_{\bar m \bar m 0} & \to & \frac{\kappa a^2 E_{\rm ISCO}}{4 \sqrt{2 \pi} M r_+^3}
\biggl[ - i \frac{k r_+}{\kappa} + \biggl(\frac{k r_+}{\kappa}\biggr)^2 \biggr]
\nonumber \\  && \times
\frac{r_+ - i a \cos \theta_0}{r_+ + i a \cos \theta_0} \sin^2 \theta_0 S_{\sub} \epsilon^{-1} \ ,
\end{eqnarray}
where $\theta_0$ is the value of $\theta$ that the point particle asymptotes to on the horizon, and $E_{\rm ISCO}$ contains information on the particle's constants of motion. This equation differs from MB by the factor of $-1$ in front of the term $ i k r_+ /\kappa$. When this difference is accounted for, we have for (MB 3.6)
\begin{equation}
\label{eq3.23}
\tilde{Z}_{\sub} \propto (\omega - m \Omega_H + i g_H)(\omega - m \Omega_H +2 i g_H) \ .
\end{equation}
The first root of $\tilde{Z}_{\sub}$ then cancels the pole in the denominator of (MB 3.7). It also appears that the second root removes the frequency mode our rough argument in Section IIIA suggests. However, \eqref{eq3.23} holds only to first order, and at second order there are additional terms not proportional to these roots. While a second surprising cancellation can only be ruled out by a careful study of the MB analysis at second order (or an equivalent formulation), it would seem unlikely that the next order of frequency mode would vanish as well. Such a careful study goes beyond the scope of this paper. However, investigation does indeed show that at second order there is a HM with a decay rate of $2 g_H$ \cite{ChenZimmerman}.

\section{Numerical Study}

\begin{figure}[tb] 
  \centering
  \includegraphics[width=3.375 in, keepaspectratio]{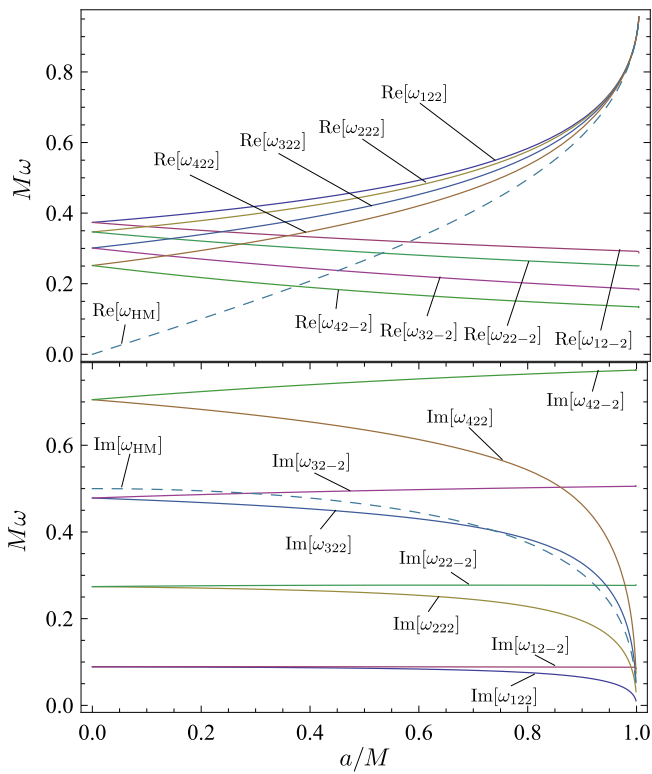}
  \caption{(color online) A comparison of the real and imaginary parts of $\omega_H$ (dashed) and the first four QNMs, with $\ell = 2 \,, m= \pm 2$ (solid).}
  \label{fig:FrequencyPlot}
\end{figure}

\begin{figure}[tb]
\includegraphics[width=3.375in, keepaspectratio]{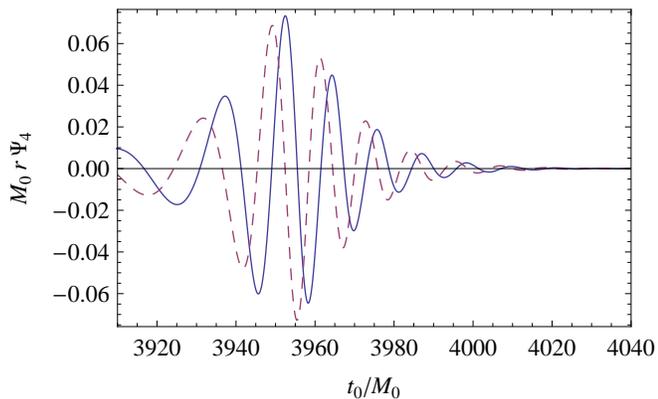}
\caption{(color online) The section of the numerical waveform given in Scheel \textit{et. al.} over which $t_0$ ranges in the overlap integral. Here we give both the real (solid) and imaginary (dashed) portions of $M_0 r \Psi_4$.}
\label{fig:OverlapWaveform}
\end{figure}

In the previous sections we argued for the presence of a HM in the ringdown spectrum. In this section we test a numerical waveform for evidence of this mode. For this study we use the publicly available waveform generated by Scheel \textit{et. al.} \cite{Scheel} by the evolution of an equal mass black hole binary through merger and ringdown. First we compute the overlap between the final portion of the numerical waveform and either a combination of QNM oscillations, or a combination of QNM oscillations and the HM, in order to see if the given combination is a good fit to the waveform. Next, we use a best-fit of the overlap to extract the mass and spin of the black hole from the waveform. Again, we compare the extraction using just the QNMs with an extraction which includes the HM.

For our HM, we focus on the less damped of the two frequencies discussed in Section II, $\omega_H = m \Omega_H - 2 i g_H$. This mode agrees with the frequency mode from the point particle plunge discussed in Section III, and also from the more physically motived model of the perturbed black hole's transition into the linear perturbation regime, discussed in Section IIC. Throughout this section we will discuss only the dominant $\ell =2$, $m=2$ mode in the spherical harmonic decomposition of the waveform. Note that, because of the rotation of the Kerr black hole, there is a Zeeman-like splitting of the QNMs into modes which corotate with the black hole and modes which counter-rotate with the hole. The counterrotating mode frequencies are equivalent to the $\ell =2, \ m=-2$ QNM frequencies, but with a negative real part of the frequency, $\omega^{\rm CR}_{n \ell m} = - \bar \omega_{n \ell -m}$ (see \cite{BertiLisaTest} for a detailed discussion). Each additional overtone we consider in the numerical analysis in the following sections thus adds two distinct modes. 

Figure \ref{fig:FrequencyPlot} compares the real and imaginary parts of $\omega_H$ with those of the first four QNM for $\ell =2$,\  $m=2$ (corotating modes) and $\ell =2, \ m=-2$ (counterrotating modes), as a function of $a/M$. These QNM frequency values are drawn from \cite{BertiComm} and calculated using the methods discussed in \cite{Berti}, whose values are used throughout this study. Note also that while many studies refer to $n=0$ as the slowest decaying QNM, here we count overtones from $n=1$. The corotating QNMs generally have a higher frequency than the HM, and the first two QNMs have slower decay rates. The third corotating QNM has a comparable decay rate, and the fourth decays faster than the HM. Meanwhile, the counterrotating QNMs decrease in frequency with increasing $a/M$, until they oscillate slower than the HM. The decay rates of the counterrotating QNMs also remain nearly constant over the whole range of  $a/M$, and so in this case the third and fourth QNMs both decay more quickly than the HM over a large range of $a/M$. As $a/M$ goes to zero, $\omega_H$ ceases to drive oscillations, and the HM simply decays exponentially. Also, we see that as $a/M$ goes to zero, the degeneracy between co and counterrotating modes is restored. Finally, in Table \ref{t:QNMs} we give numerical values for the QNMs, evaluated at $a_f/M_f$ for the final, merged black hole whose spectrum we study \cite{Scheel} (in units of the initial ADM mass of the binary), as well as the value of the HM at this spin parameter. 

\begin{table}[t]%
\label{t:QNMs}
\centering %
\begin{tabular}{cc}
\toprule %
 $(n, \ \ell,\  m)$  & $M \omega_{\rm QNM}$  \\ \hline
(1,2,2)& $ 0.52670 +  0.08129 i$ \\
(2,2,2)  &$0.51486 + i 0.24581 i$\\
(3,2,2)  &$0.49296 +  0.41513 i$\\
(4,2,2)  &$0.46387 +  0.58873 i$\\
(5,2,2)  &$0.43291 +  0.76035 i$\\
(1,2,-2) &$0.31072 +  0.08874 i$ \\
(2,2,-2) &$0.27312 +  0.27733 i$\\
(3,2,-2) &$0.21198 +  0.49963 i$\\
(4,2,-2) &$0.15865 +  0.75811 i$\\
(5,2,-2) &$0.12707 +  1.03031 i$\\
 \toprule
 $m $ & $M \omega_{\rm HM}$\\ \hline
$2$ &  $0.37177 +  0.43089 i$
\end{tabular}
\caption {Values of the QNMs and HM evaluated at the spin parameter $a_f /M_f = 0.68646$ appropriate for the final black hole of \protect\cite{Scheel}.}
\end{table}

\subsection{Comparison of Overlaps}

\begin{figure}[tb]
\includegraphics[width=3.375in, keepaspectratio]{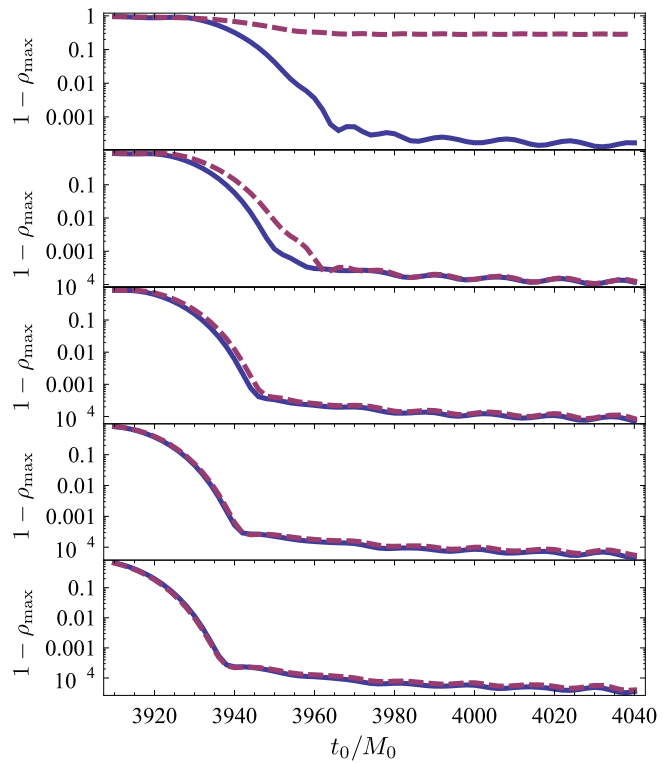}
\caption{(color online) Comparisons of $1-\rho_{\rm max}$ for overlaps using just the QNMs (solid) and those which replace one QNM pair with the HM (dashed). The top panel compares the first QNM pair (both co and counterrotating) with the HM alone. The second panel compares the first two QNM pairs with the first QNM pair and the HM. Subsequent panels compare the first $n$ ($n = 3, 4, 5$ )QNM pairs with the first $n-1$ pairs and the HM.}
\label{fig:DiffOverlaps}
\end{figure}

Consider the final portion of the numerical waveform, $\psi(t) H(t-t_0)$, with $H(t)$ the unit step function, and $t_0$ a constant which we consider to be the time where the ringdown phase begins. We wish to see how well a waveform made from a linear combination of damped sinusoids, $\psi_k = e^{(-i \omega_k -\gamma_k)(t-t_0)} H(t-t_0)$, can be made to fit $\psi$.  Since we do not know {\it a priori} at what point in the numerical waveform the ringdown phase begins, we vary $t_0$ as a free parameter in our study. This allows us to see where in the waveform our combination of sinusoids fails to be a good fit; at sufficiently early $t_0$ we do not expect a particular combination of $\psi_k$ to model the chosen section of the waveform accurately. However, a combination of $\psi_k$ that fits the waveform well over a range of $t_0$ that includes the early parts of the ringdown more accurately represents the frequency spectrum of the ringdown than another set of damped sinusoids that first fails at a larger value of $t_0$.

For two waveforms $s_1(t)$ and $s_2(t)$ cut off at $t_0$, we first define the inner product
\begin{equation}
\label{eq4.1}
\langle s_1 | s_2\rangle \equiv \int_{t_0}^{+\infty} \bar{s}_1(t) s_2(t) dt \ .
\end{equation}
The overlap, $\rho$, of two waveforms is given by the magnitude of the normalized inner product of the waveforms. Our goal then is to maximize the overlap of $\psi(t_r)$ and the combination $\sum\limits_k \alpha_k \psi_k$,
\begin{equation}
\label{eq4.2}
\rho^2 = 
 \frac{|\big \langle   \sum\limits_{k} \alpha_k \psi_k| \psi\big\rangle|^2}{{\langle \sum\limits_k\alpha_k \psi_k|\sum\limits_k\alpha_k \psi_k } \rangle} = \frac{\sum\limits_{k,j} \bar{\alpha}_k A_k \bar{A}_j \alpha_j}{\sum\limits_{l,m} \bar{\alpha}_l B_{lm}\alpha_m} \ ,
\end{equation}
where $A_k \equiv \langle \psi_k|\psi \rangle$ , $B_{l m} \equiv \langle \psi_l| \psi_m\rangle$. This maximum overlap characterizes how well the $\psi_k$ can be made to approximate $\psi$, given the optimum choice of $\alpha_k$. The maximization yields\footnote{The Lagrange multiplier method yields $\vec A \vec A^\dagger \vec\alpha = \lambda \mathbf{B} \vec\alpha$, with $\lambda$ already the extremum.  This means $\lambda$ should be the maximum eigenvalue of $\mathbf{M} = \mathbf{B}^{-1}\vec A\vec A^\dagger$. However, since $\mathbf{M}$ only has one non-zero eigenvalue, we have $\lambda =\mathrm{tr}\mathbf{M} = \vec A^\dagger \mathbf{B}^{-1}\vec A$.}

\begin{equation}
\label{eq4.3}
\rho^2_{\rm max} [\psi ;\{\omega_k,\gamma_k\}] = \frac{\bar{A}_k B^{-1}_{kj} A_j}{\langle \psi | \psi \rangle} \ .
\end{equation}

Using this equation to compute $\rho_{\rm max}$, we take the first $n$ QNMs (recall that each overtone includes two frequency modes, both the co and counterrotating modes) and compute the maximum overlap with the numerical waveform as a function of the starting time $t_0$. We then find the maximum overlap using the first $(n-1)$ QNMs and the HM. Here, and throughout this section, we normalize our units by $M_0$, the sum of the initial ADM masses for the two black holes that merge \cite{Scheel}. In these units, the mass $M_f$ of the final black hole is given as $M_f/M_0 = 0.95162 \pm 0.00002$,  and its spin $a_f$ by $a_f/M_0 = 0.65325\pm 0.00004$. For large values of $t_0$, only the least damped mode contributes to the waveform, and so any $\psi_k$ that includes the first QNM will provide $\rho_{\rm max} \approx 1$. Therefore it is useful to investigate $1-\rho_{\rm max}$. Figure \ref{fig:OverlapWaveform} plots the segment of the waveform over which we range $t_0$ for the overlap calculations. Figure \ref{fig:DiffOverlaps} compares $1- \rho_{\rm max}$ on a log scale for overlaps using the QNMs alone to those including the HM. In the context of gravitational wave signals, it is preferable to have a good overlap for $t_0$ close to the peak of the signal. This is when the gravitational wave signal is strongest, and also the point in the waveform when the number of QNMs that make a significant contribution to the waveform is the greatest, before the most-damped QNMs become negligible.

We see that the horizon mode alone provides a poor fit for the waveform for all values of $t_0$, and that the first two QNMs provide a better fit at earlier $t_0$ than the first QNM and the horizon mode. Replacing the $n^{\rm th}$ QNM with the HM gives comparable results at $n=4$, and provides a (very slight) improvement at early $t_0$ for $ n\geq 5$. We note that this improvement becomes apparent at a $t_0$ earlier than the peak value of $|\psi|$, which occurs at $t_{\rm max} = 3953.8 M_0$. This means even for $n \geq 5$, the HM makes an improvement only when portions of the waveform which should not  be well modeled by a set of damped sinusoids are included in the overlap. In fact, we find empirically that adding modes with low decay rates always tends to improve overlap calculations at early values of $t_0$. This is due to the fact that less damped modes will better fit the region near the peak of the waveform. In this case we would expect comparable results from the $n=3$ mode addition and the HM mode addition, and for the HM to improve the overlap compared to $n=4$; however, the single HM must compete with the pair of co and counterrotating modes that make up the next overtone. For these reasons, we find that overlap comparisons do not provide a compelling case for the presence of the HM in the waveform, nor do they rule the mode out.

As a second test, we would like to investigate the use of the HM in performing parameter extraction from a ringdown waveform. This practical test of the utility of the HM is more physically motivated than overlap comparisons, and can provide better evidence for the presence of the HM in the waveforms.

\subsection{Extraction of Mass and Black Hole Spin}

As a second test, we will extract the mass $M$ and the spin parameter $a$ from the waveform. We extract the mass first. To do this, we set $a/M_0$ to the value given in \cite{Scheel} for the final black hole, but allow $M$ to vary (note that the QNMs and the HM are function of $a/M$, not $a/M_0$).  We then calculate $\rho_{\rm max}$ as a function of $M$, using equation (\ref{eq4.3}), and find the value of $M$ which maximizes $\rho_{\rm max}$. When doing so, there is a distinct residual oscillation in the extracted value of $M/M_0$. Investigation reveals that the residual oscillation is compensated for by including the first $\ell =4$, $m=4$ corotating QNM in the fit. Appendix C gives a brief discussion of the possible sources of this mode mixing in the numerical waveform. 

\begin{figure}[htp]
  \includegraphics[width=3.375in, keepaspectratio]{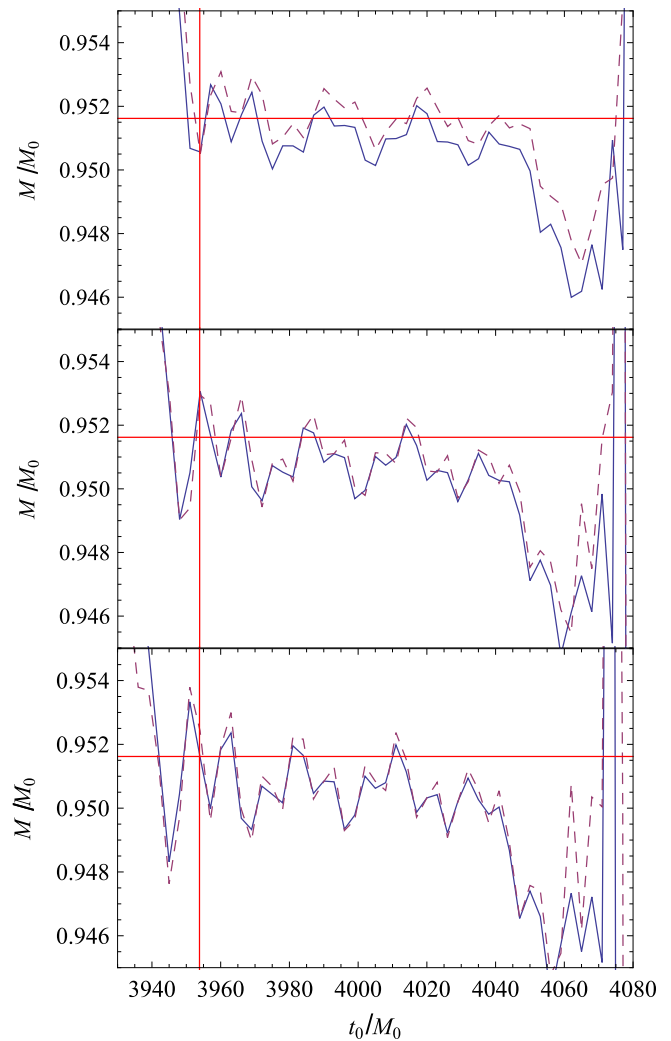}
  \caption{(color online) Values for the extracted masses $M_{\rm QNM}$ (solid) and $M_{\rm HM}$ (dashed) as functions of $t_0$, for three extractions. The top panel compares the first three QNM pairs with the first two and the HM. The middle panel compares the first four QNM pairs with the first three and the HM. The bottom panel compares the first five QNM pairs with the first four and the HM. The solid horizontal line is at $M/M_0 = 0.95162$, the mass of the final black hole as given in \protect\cite{Scheel}. The solid vertical line gives the peak of $|\psi|$.}
  \label{fig:Mextract}
\end{figure}

The top panel of Figure \ref{fig:Mextract} gives the extracted $M$ as a function of $t_0$, for two sets of frequency modes. The first set is composed of the first three $\ell = 2$, $m=2$ QNM pairs (co and counterrotating), plus the first corotating $\ell=4$, $m=4$ QNM. The second set replaces the third $\ell=2$, $m=2$ QNM pair with the HM.  We expect the extraction to fail at early values of $t_0$, where the sinusoids are a poor fit to the waveform, and at late values of $t_0$, where the waveform has decayed significantly. Indeed, one can see in the top panel of Figure \ref{fig:Mextract} that the extraction begins to diverge as portions of the waveform preceding the peak of $|\psi|$ at $t_{\rm max} = 3953.8 M_0$ (shown with a solid vertical line) are included in the extraction, corresponding to values of $t_0$ earlier than the peak. The middle and bottom panels of Figure \ref{fig:Mextract} show the results of the same extraction, using the first four and five $\ell = 2, \ m=2$ QNM pairs, respectively, and comparing to extractions which replace the QNM pair with the largest decay with the HM. We see in these figures that the extraction can be carried out to even earlier values of $t_0$ than the peak of $\psi$, but since these extractions include portions of the waveform which do not correspond to ringdown, we do not expect these early time extractions to be accurate. 

In all cases, Figure \ref{fig:Mextract} shows that the substitution of the HM does not improve the extraction over the next most-damped QNM. We find the mean and RMS deviation of the extraction over an interval $t_0/M_0 = [3954, 4074]$ for all three extractions. This interval covers a region of $t_0$ that begins just outside the peak and continues until the extractions begin to diverge rapidly. We compare to the mass given in \cite{Scheel} giving a difference in extracted mass of $\Delta M_{\rm QNM}= M_f - M_{\rm QNM}$ for the fits that use only QNMs and  $\Delta M_{\rm HM}= M_f - M_{\rm HM}$ for those that include the HM. We present the values of $\Delta M$ in Table I. We see that in all cases, the HM gives a comparable extraction. For the extractions that use a larger number of QNMs, we see that the RMS deviation grows. This appears to be due to the fact that the extractions with a larger number of modes diverge slightly earlier than those with $n=3$ and $n=2$ QNMs shown in the top panel of Figure \ref{fig:Mextract}.

\begin{table}[t]%
\label {t:Extractedtable}
\centering %
\begin{tabular}{ccc}
\toprule %
 $n$ QNMs & $\Delta M_{\rm QNM}$ & $\Delta a_{\rm QNM}$\\
   & $( M_0 \times 10^{-3} )$ & $( M_0 \times 10^{-3} )$ \\ \hline
3& $1.23\pm 1.66$ & $0.20 \pm 3.36$ \\
4  &$1.70 \pm 1.96 $ & $ 1.23 \pm 4.95$ \\
5 & $0.84 \pm 8.14$ & $2.55 \pm 8.74$\\ \toprule
 $n$ QNMs & $\Delta M_{\rm HM}$ & $\Delta a_{\rm HM}$\\
   & $( M_0 \times 10^{-3} )$ & $( M_0 \times 10^{-3} )$ \\ \hline
2& $0.53\pm 1.37$ & $0.42 \pm 2.95$ \\
3  &$1.15 \pm 1.73 $ & $ 1.57 \pm 5.07$ \\
4 & $0.40 \pm 8.03$ & $3.00 \pm 8.93$\\ 
\end{tabular}
\caption {Extracted masses and spin parameters, for extractions using the first $n$ QNMs, and extractions using the first $n$ QNMs plus the HM.}
\end{table}

\begin{figure}[htp] 
  \includegraphics[width=3.375in, keepaspectratio]{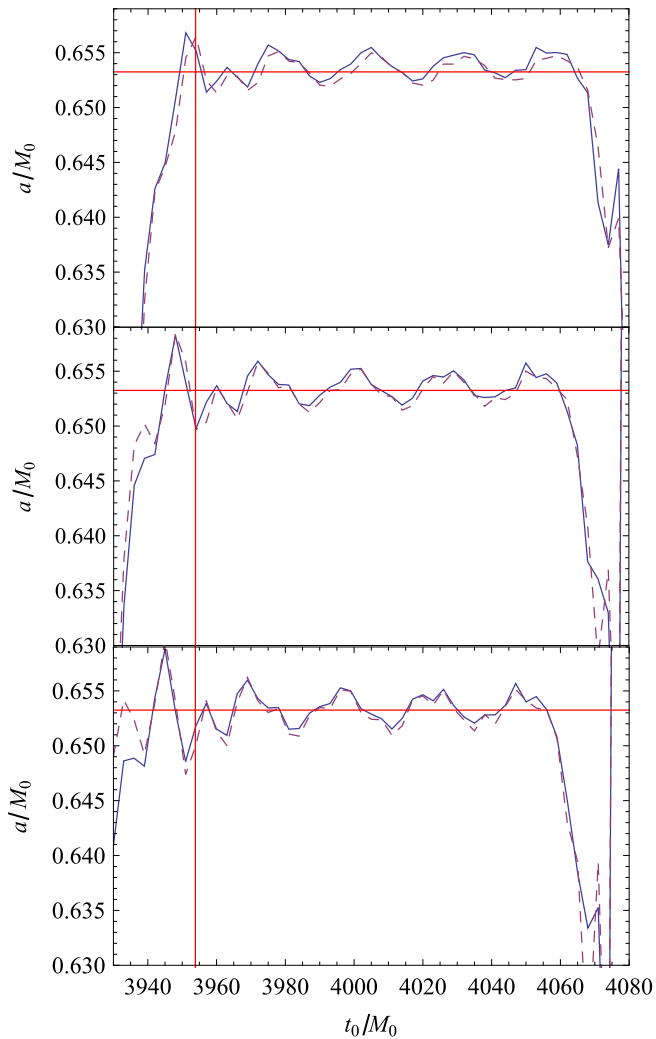}
  \caption{(color online) Values for the extracted spins $a_{\rm QNM}$ and $a_{\rm HM}$ as functions of $t_0$, for three extractions. The top panel compares the first three QNM pairs with the first two and the HM. The middle panel compares the first four QNM pairs with the first three and the HM. The bottom panel compares the first five QNM pairs with the first four and the HM. The solid horizontal line is at $a/M_0 = 0.65325$, the spin of the final black hole as given in \protect\cite{Scheel}. The solid vertical line gives the peak of $|\psi|$.}
  \label{fig:aExtract}
\end{figure}

We carry out the same test for the spin of the black hole $a/M_0$, by fixing the mass $M$ at its final value and allowing $a/M_0$ to vary. The results are given in Figure \ref{fig:aExtract}. The  situation is the same as in the case of the mass extractions. We calculate the mean and RMS deviation from the mean on the interval $t_0/M_0 = [3954, 4074]$, and compare the extracted spin parameters to that given in \cite{Scheel}, and give the results for $\Delta a$ in Table \ref{t:Extractedtable}. Once more, extractions with just QNMs are essentially the same as those with the HM replacing the most rapidly decaying QNM of a given set. From these extraction tests, we cannot conclude that the HM is present in the numerical waveforms.

\section{Discussion}

Using two methods, we have found an additional ringdown mode for the Kerr black hole. This HM depends only on the fundamental properties of the black hole: it oscillates at the horizon frequency of the black hole, and decays at a rate proportional to the surface gravity of the black hole. It will arise when generic initial perturbations source linear gravitational radiation, a situation that would occur as the spacetime transitions from a regime of stronger, nonlinear perturbations into a final ringdown phase. This occurs at the last stage of a compact binary merger, or stellar core collapse resulting in a black hole. We emphasize that this mode is not in the QNM spectrum which is generally taken as the complete spectrum for the ringdown of a black hole. At the same time, this oscillation mode is part of what is normally considered the ``ringdown'' phase of an event that results in a  final black hole, since it arises in linearized perturbation theory about the final black hole.

In fact we have discussed two possible decay rates for the HM, each dependent on our model of how the spacetime transitions from nonlinear evolution into the linear regime. One mode, found using a naive model of transition at a set time slice, decays rapidly. The second was found by noting that nonperturbative regions of the spacetime should be bounded by in-going and out-going characteristics, and is physically better motivated. It has a decay rate $\gamma = 2 g_H$, approximately the same decay rate as the $n = 3, \ell = 2, m=2$ QNM. We find that this mode has the same influence on overlap calculations as the $n=4$ pair of QNMs, though it does not appear to improve parameter extraction over the use of an additional QNM pair with $n \geq 4$. Due to its comparable decay rate to the $n=3$ QNMs,  it should be considered in the construction of waveform templates that use $n \geq 3$ QNMs. The HM should also be included as part of the ringdown spectrum when considering the potential use of an observed ringdown signal as a test of the No-Hair Theorem. Otherwise, the presence of this non-QNM oscillation in the spectrum might lead one to conclude that the signal was emitted from an object other than a Kerr black hole.

The analytic approach presented here also builds some intuition into the origin of various frequency modes of linear perturbations of the Kerr spacetime. The HM studied here arises when the influence of perturbations near the horizon are considered. Integration of these initial perturbations using the Green's function approach results in the presence of a pole in the frequency integral of equation \eqref{eq2.8}. This mode depends only on the properties of the black hole which govern its near horizon geometry. Meanwhile, the usual QNMs arise because of the poles in the Wronskian of the radial Green's function, equations \eqref{eq2.20}-\eqref{eq2.21}. In our model, these modes arise due to the interaction of the initial perturbations with the complicated potential of the wave equation present  further from the event horizon, a situation analogous to that explored for Schwarzschild black holes by Price \cite{Price}. In this work, decaying perturbations on the surface of a collapsing star are associated with out-going radiation; comparison of our results with \cite{Price} indicates that our HM is associated with the decaying mode at the stellar surface, but that the rotation of the Kerr black hole in our case guarantees that this mode oscillates with the horizon frequency in addition to its simple decay.

We have also reviewed the problem of gravitational radiation from a point particle infalling near the horizon. Previous work \cite{MinoBrink} both motivated this study and guided our investigation. However, our results conflict with those of the motivating study. In investigating this discrepancy, we have found an error in the original calculation of \cite{MinoBrink}, the correction of which cancels the first order results for the radiation at infinity. We have also argued that the form of the next order correction agrees with our results for vacuum perturbations. We leave the detailed calculation of the correct second order terms to a future study \cite{ChenZimmerman}.

Future study using a variety of numerical waveforms will be key in determining the importance of the HM in template generation and gravitational wave detection. In a simulation where the excitation of slowly decaying QNMs is suppressed, we would expect the HM to be a clear component of the ringdown. Future study of how one might suppress this QNM excitation would be valuable, and such simulations would provide the best testing ground for the presence of the HM in numerical simulations. In addition, the properties of the near horizon region, the HM itself, and the regularity conditions on the initial data discussed here may be of interest in the mathematical study of the stability of the Kerr black hole (see e.g. \cite{Stability} and the references therein).

\section{Acknowledgments}

We gratefully thank E. Berti for providing tables of QNM values, and also for valuable discussions about the counterrotating QNMs of the Kerr black hole. We thank Jeandrew Brink, Jim Isenberg, David Nichols, and Kip Thorne for useful discussions. A. Z. would also like to thank the National Institute for Theoretical Physics of South Africa for hosting him during the completion of this work. This work was supported by NSF Grants No. PHY-0601459, No. PHY-0653653, No. PHY-1068881, CAREER Grant No. PHY-0956189, the David and Barbara Groce Startup Fund at Caltech, and the Brinson Foundation.

\appendix

\section{Green's Function Formalism For the Teukolsky Equation}

The Teukolsky equation, for spin $s = -2$, can be written in Boyer-Lindquist coordinates using the Kinnserly tetrad as \cite{Teukolsky}
\begin{equation}
\label{eqa.1}
L[\psi] = \Delta^{-2} T \ ,
\end{equation}
with $L$ the linear Teukolsky operator described below; $\psi = \Psi_4/\rho^4$; and the source term $T$ a complicated function of the stress-energy tensor $T^{\mu \nu}$, the Kinnersley tetrad, and the Kerr rotation coefficients. The Teukolsky operator is
\begin{eqnarray}
\label{eqa.2}
L &=&  L_r + L_{\theta} + A_1 \partial_t^2 +  A_2 \partial_t+ A_3 \partial_t \partial_{\phi} +A_4 \partial_{\phi}^2 
\nonumber \\ & &
+ A_5 \partial_{\phi} + A_6 \ , \\
\label{eqa.3}
L_r & = &  - \partial_r( \Delta^{-1} \partial_r) \ , \\
\label{eqa.4}
L_{\theta} & = &- \frac{1}{\Delta^2 \sin \theta} \partial_{\theta} (\sin \theta \partial_{\theta}) \  , \\
\label{eqa.5}
A_1 & = &  \frac{(r^2+a^2)^2}{\Delta^3} - \frac{a^2 \sin^2 \theta}{\Delta^2} \ , \\
\label{eqa.6}
A_2 & = & \frac{4M(r^2 -a ^2)}{\Delta^3} - \frac{4(r + i a \cos \theta)}{\Delta^2} \ , \\
\label{eqa.7}
A_3 & = & \frac{4 M a r}{\Delta^3} \ , \\
\label{eqa.8}
A_4 & =& \frac{a^2}{\Delta^3} - \frac{1}{\Delta^2 \sin^2 \theta} \ , \\
\label{eqa.9}
A_5 & = & \frac{4a(r-M)}{\Delta^3} + \frac{ i \cos \theta}{\Delta^2 \sin^2 \theta} \ , \\
\label{eqa.10}
A_6 & =&  \frac{4 \cot^2 \theta + 2}{\Delta^2} \ .
\end{eqnarray}

We introduce the adjoint operator $L^*$, which is the Teukolsky operator with the substitutions $(\partial_t \to -\partial_t, \partial_{\phi} \to -\partial_{\phi})$, and the Green's function $G(x'^\mu; x^{\mu})$ for $L$, which obeys $L[G(x'^{\mu};x^{\mu})] = L^*[G(x'^{\mu};x^{\mu})] = \delta(t' -t) \delta(r'-r) \delta(\theta' - \theta) \delta (\phi' - \phi) \equiv \delta^4(x'^{\mu} - x^{\mu})$. Now, note that given a pair of functions $u$ and $v$ we have
\begin{eqnarray}
\label{eqa.11}
&& u L^*[v] - v L[u] = \partial_r [ \Delta^{-1} (v \partial_r u - u \partial_r v)] 
\nonumber \\  &&
\qquad + \frac{1}{\sin \theta}\partial_{\theta}[ v \sin \theta \partial_{\theta} u - u \sin \theta \partial_{\theta}v]
\nonumber \\  &&
\qquad + A_1 \partial_t [ u \partial_t v - v \partial_t u] 
 - A_2 \partial_t [uv]  
\nonumber \\  &&
\qquad + A_3 [ \partial_t( u \partial_{\phi} v) - \partial_{\phi}(v \partial_t u)]
 + A_4 \partial_{\phi}[u \partial_{\phi} v - v \partial_{\phi} u]
\nonumber \\  &&
\qquad -A_5 \partial_{\phi}[uv] \ .
\end{eqnarray}

Now, we let $u = \psi(x^{\mu})$, $v = G(x'^{\mu};x^{\mu})$, and we integrate the entire expression over the domain of interest for our situation, $t \in [0, \infty)$, $r \in (r_+, \infty)$, $\theta \in [0, \pi]$, $\phi \in [0, 2\pi]$, using spherical polar coordinates and a Euclidean volume element $d^4x \equiv dt d^3 x =  r^2 \sin \theta\ dt dr d\theta d \phi$. To evaluate the left side of equation (\ref{eqa.11}) we note
\begin{eqnarray}
\label{eqa.12}
 \int d^4 x \  \psi(x^{\mu}) L^*[G(x'^{\mu};x^{\mu})] 
&=&  \int d^4 x \  \psi(x^{\mu}) \delta^4(x'^{\mu} - x^{\mu})
\nonumber \\ 
& = & \psi(x'^{\mu}) \ .
\end{eqnarray}
Also, we have
\begin{eqnarray}
\label{eqa.13}
\int d^4 x \ G(x'^{\mu};x^{\mu}) L[\psi(x^{\mu}) ] 
& = & \int d^4 x \ G(x'^{\mu};x^{\mu})  \Delta^{-2} T \ .
\nonumber \\
\end{eqnarray} 

On the right hand side of equation \eqref{eqa.11}, we note that the terms involving $A_4$ and $A_5$ vanish when integrated over $\phi$, due to the periodicity of $\phi$. The term $- A_3 \partial_{\phi}(v \partial_t u)$ vanishes for the same reason. The first term becomes a boundary term when integrated over $r$. In order to have only in-going waves at the horizon, and only out-going waves at infinity, we must impose homogeneous boundary conditions on $\psi(x^{\mu})$ and $G(x'^{\mu};x^{\mu})$, and so this term also vanishes. The term involving derivatives of $\theta$ vanishes when integrated over $\theta$, since we require that the initial data and the Green's function be regular on the boundary of  $[0, \pi]$. The terms involving $A_1$ and $A_2$ are total derivatives in time, and so when we integrate over $t$ we remove the time derivatives and evaluate the terms on the boundary at $t = 0$. Since our physical source is transient, the terms vanish at the bound of $t \to \infty$. We have
\begin{eqnarray}
\label{eqa.14}
\psi(x'^{\mu}) & =&  \int d^4 x \ G(x'^{\mu};x^{\mu})  \Delta^{-2} T 
\nonumber \\ &&
+ \int d^3 x \  A_1 \partial_{t} G(x'^{\mu};x^{\mu}) \psi(x^{\mu}) \bigg|_{t=0}
\nonumber \\ &&
- \int d^3 x \  A_1  G(x'^{\mu};x^{\mu})\partial_{t} \psi(x^{\mu})\bigg|_{t=0}
\nonumber \\ & &
-\int d^3x \ A_2  G(x'^{\mu};x^{\mu}) \psi(x^{\mu})\bigg|_{t=0} 
\nonumber \\ & &
+\int d^3x \  A_3 \partial_{\phi} G(x'^{\mu};x^{\mu}) \psi(x^{\mu})\bigg|_{t=0}
\end{eqnarray}
Thus far we have kept the source term $T$ in place for comparison with other studies of the evolution of the Teukolsky equation. In this study we are interested in the vacuum case, and so we set $T=0$ here and throughout Section II.

We can further simplify this expression for $\psi(x^{\mu})$ by performing the angular integrations. Let us expand the initial perturbation in terms of spherical harmonics,
\begin{eqnarray}
\label{eqa.15}
\psi(t , r, \theta, \phi) \bigg|_{t=0} &=& \sum_{\ell',m'} a_{\ell' m'} (r) Y_{\ell' m'} (\theta, \phi) \  , \\
\label{eqa.16}
\partial_t \psi(t , r, \theta, \phi) \bigg|_{t=0} & = & \sum_{\ell',m'} b_{\ell' m'} (r) Y_{\ell' m'} (\theta, \phi) \  .
\end{eqnarray}
In addition, we expand the Green's function in the frequency domain, where it can be written down explicitly in terms of the spin-weighted spheroidal harmonics and the Green's function for the radial Teukolsky equation \cite{Teukolsky,FrolovNovikov},
\begin{eqnarray}
\label{eqa.17}
G(x'^{\mu};x^{\mu}) \bigg|_{t=0} &=& \int \frac{d\omega} {(2 \pi)^2} e^{-i \omega t'} 
\sum_{\ell,m} \tilde{G}_{\sub} (r',r) 
\nonumber \\ & & \times
S_{\sub}(\theta') \bar{S}_{\sub}(\theta) e^{i m (\phi' - \phi)} \ .
\nonumber \\
\end{eqnarray}
So, we have $\partial_{\phi} G(x'^{\mu};x^{\mu}) = - i m G(x'^{\mu};x^{\mu})$. With this, we can now perform the integration over $\phi$, using the identity
\begin{equation}
\label{eqa.18}
\int_0^{2\pi} \frac{d\phi}{2 \pi} e^{i (m-m') \phi} = \delta_{m m'} \  ,
\end{equation}
which allows us to resolve the summation over $m'$ contained in equation \eqref{eqa.15}-\eqref{eqa.16}. From here, it is convenient to impose the near horizon approximation, for which the motivation is discussed in Section IIB. We keep terms only to the leading order in $\epsilon = (r-r_+)/r_+ \ll 1$. In this approximation, we have that $\Delta \approx 2Mr_+\kappa\epsilon$, with $\kappa \equiv \sqrt{1 - a^2/M^2}$. To first order in $\epsilon$,
\begin{eqnarray}
\label{eqa.19}
A_1 & \approx & (2 M r_+)^{-1} (\kappa \epsilon)^{-3} \ , \\
\label{eqa.20}
-i m A_3 - A_2  & \approx & - \frac{2 M \kappa + i m a}{2(M r_+)^2 (\kappa \epsilon)^3} \ . 
\end{eqnarray}
We note that all $\theta$ dependence for these functions enters in at second order in the near horizon expansion. We define 
\begin{eqnarray}
\label{eqa.21}
\alpha_{\sub} (r) & \equiv & \sum_{\ell'} a_{\ell' m}(r)\sqrt{\frac{(2 \ell' +1)(\ell'-m)!}{4 \pi (\ell'+m)!}} 
\nonumber \\ & & \times
\int_0^{\pi} \sin \theta d\theta   P_{\ell' m}(\cos\theta) \bar{S}_{\sub}(\theta) ,
\\
\label{eqa.22}
\beta_{\sub} (r) & \equiv & \sum_{\ell'} b_{\ell' m}(r)\sqrt{\frac{(2 \ell' +1)(\ell'-m)!}{4 \pi (\ell'+m)!}} 
\nonumber \\ & & \times
\int_0^{\pi} \sin \theta d\theta   P_{\ell' m}(\cos\theta) \bar{S}_{\sub}(\theta) ,
\end{eqnarray}
where $P_{\ell m} (x)$ are the associated Legendre polynomials. The functions $\alpha_{\sub}(r)$ and $ \beta_{\sub}(r)$ are nonzero only on the interval $r \in [r_+, (1+\xi)r_+]$, which allows us to truncate the radial integrals in equation \eqref{eqa.14}. In fact, we only desire the leading order behavior in $\epsilon$ of these functions, and this is discussed in Section IIB.

Inserting equations \eqref{eqa.15} - \eqref{eqa.21} into \eqref{eqa.14}, and exchanging primed and unprimed labels, we have finally
\begin{eqnarray}
\label{eqa.23}
\psi(x^{\mu}) & = & \int \frac{d \omega}{2 \pi}  \sum_{\ell m} e^{-i \omega t + i m \phi}R_{\sub} (r) S_{\sub}(\theta) , \\
\label{eqa.24} 
R_{\sub} (r) & = & - \int_{r_+}^{(1+\xi)r_+} dr' \biggl[ \frac{\beta_{\sub}(r') + i \omega \alpha_{\sub}(r') }{2 M r_+ (\kappa \epsilon)^3} 
\nonumber \\ && \qquad 
+ \frac{(2 M \kappa + i m a) \alpha_{\sub}(r')}{2(M r_+)^2 (\kappa \epsilon)^3} \biggr] \tilde{G}_{\sub}(r,r') \ .
\nonumber \\
\end{eqnarray}
We resolve this expression in Section IIB.

\section{The Teukolsky Equation in the Newman-Penrose Formalism}

We refer the reader to \cite{FrolovNovikov, NewmanPenrose} for the full formalism. Here we simply collect some of the longer expressions used for Section IIIA.

From (2.14) of \cite{Teukolsky} we have
\begin{eqnarray}
\label{eqb.1}
&&\Big[(\hat\Delta +3\gamma-\bar\gamma+4\mu+\bar\mu)(\hat D+4\epsilon-\rho)  \nonumber\\
&&-(\bar\delta-\bar\tau+\bar\beta+3\alpha+4\pi)(\delta-\tau+4\beta)\nonumber\\
&&\qquad\qquad\qquad\qquad\qquad\qquad\; -3\Psi_2\Big]\Psi_4^B = 4\pi T_4.\quad\quad
\end{eqnarray}
Here $\Psi_2$ refers to the background value of the NP scalar, $\Psi_2 = M \rho^3$ for Kerr. The scalar $\Psi_4^B$ is the perturbative value of $\Psi_4$, which is zero at leading order for Kerr. Here, $\hat{D},\  \hat{\Delta},\  \delta$ are all derivative operators along the directions of the null basis, and the Greek characters represent combinations of the spin coefficients. Also note the unfortunate but standard use of $\pi$ on the left-hand side to refer to one of the spin coefficients in the null tetrad, while on the right side it refers to the numerical $\pi$ from the Einstein field equations. It is generally clear which is which, and in any case the NP coefficient enters at subleading order here. The source term $T_4$ is given by
\begin{eqnarray}
\label{eqb.2}
T_4 &=& (\hat\Delta+3\gamma-\bar\gamma+4\mu+\bar\mu)\Big[(\bar\delta-2\bar\tau+2\alpha)T_{n\bar m } \nonumber\\
&&\qquad\qquad\qquad\quad -(\hat\Delta+2\gamma-2\bar\gamma+\bar\mu)T_{\bar m \bar m}\Big] \nonumber\\
&&+(\bar\delta-\bar\tau+\bar\beta+3\alpha+4\pi)\Big[(\hat\Delta+2\gamma+2\bar\mu)T_{n\bar m} \nonumber\\
&&\qquad\qquad\qquad\quad -(\bar\delta-\bar\tau+2\bar\beta+2\alpha)T_{nn} \Big]\,.
\end{eqnarray}
Here, the terms $T_{ab}$ are the components of the stress-energy tensor in the tetrad basis, $T_{nn} = T_{\mu \nu} n^{\mu} n^{\nu}$, $T_{n \bar m} = T_{\mu \nu} n^{\mu} \bar m^{\nu}$, etc.

To specialize to the near horizon approximation, we note that $\hat{D}$ contains $\Delta^{-1}$, and therefore dominates over all the other terms.  In addition, we have $\gamma=\bar\gamma = \rho \bar \rho (r - M)/2$ and $\mu=0$, to first order.

Using the commutation relation between $D$ and $\hat \Delta$ (NP4.4), we have, near the horizon
\begin{equation}
\label{eqb.3}
\hat\Delta \hat D - \hat D \hat \Delta = 2\gamma \hat D +\mbox{(lower order terms)}
\end{equation}
which subsequently gives the $O(\Delta^{-1})$ term on the left-hand side of the equation
\begin{equation}
\label{eqb.4}
(\hat D \Delta +4\gamma\hat D) \Psi_4
\end{equation}
We investigate this expression more fully in Section IIIA.

\section{Mode Corrections to the Wavefunction}

\begin{figure}[tbh]
  \centering
  \includegraphics[width=3.375in,keepaspectratio]{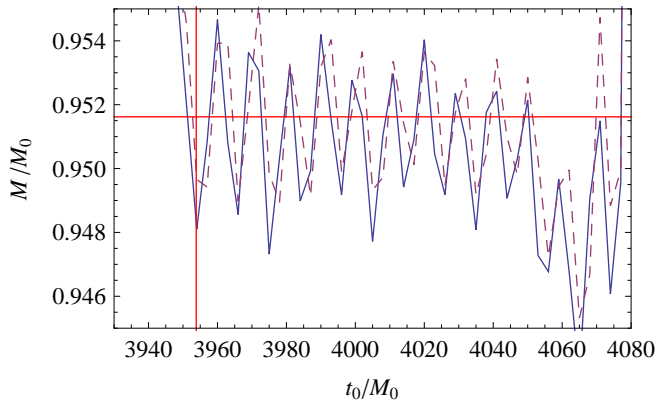}
  \caption{(color online) Extraction of the mass $M/M_0$ as a function of $t_0$, using only first three QNM pairs (solid) or the first two QNM pairs and the HM (dashed). Here we do not include the corotating $n=1,\  \ell=4,\  m=4$ mode.}
  \label{fig:MExtractNocorrect}
\end{figure}

\begin{figure}[tbh]
  \centering
  \includegraphics[width=3.375in,keepaspectratio]{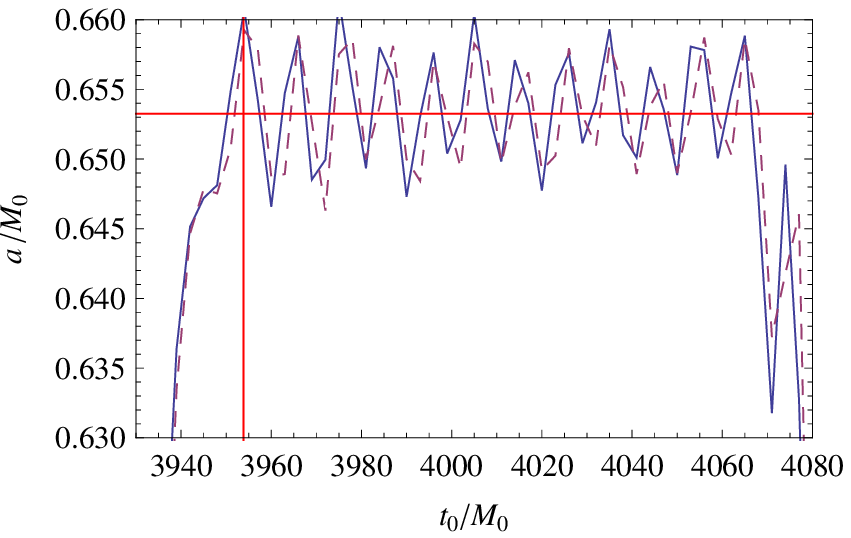}
  \caption{(color online) Extraction of the spin $a/M_0$ as a function of $t_0$, using only first three pairs QNMs (solid) or the first two QNM pairss and the HM (dashed). We do not include the corotating $n=1,\  \ell=4, \ m=4$ mode.}
  \label{fig:aExtractNocorrect}
\end{figure}

The presence of oscillation at the $n=1,\ \ell = 4, \  m=4$ corotating QNM merits some brief discussion. Figures \ref{fig:MExtractNocorrect} and \ref{fig:aExtractNocorrect} give the extraction of $M$ and $a$ using the first three QNM pairs (and comparing the the first two QNM pairs with the HM), without the $\ell = 4, \  m = 4$ mode included. Comparison with the topmost panels of Figures \ref{fig:Mextract} and \ref{fig:aExtract} shows that the distinct oscillation is successfully removed by including this mode. 

A certain amount of mode mixing between the QNMs is expected due to the fact that the waveform is decomposed into spin-weighted spherical harmonics during the extraction of the waveform. In fact, the angular eigenfunctions of the Teukolsky equation are the spin-weighted spheroidal harmonics. These functions become the usual spherical harmonics when $a \omega = 0$. Using this fact, the spheroidal harmonics can be expanded in terms of spin-weighted spherical harmonics and powers of $a \omega$, as first discussed in \cite{TeukolskyPress1}. Only spherical harmonics with the same $s$ and $m$ contribute in the expansion. As such, we see immediately that the mixing with the $\ell =4,\  m=4$ QNM frequency cannot arise from the decomposition into spherical harmonics. The portions of the waveforms that can mix into the $\ell =2,\  m=2$ waveform arise from the expansions of $S_{3 2}, S_{4 2}$, etc. Explicitly, the expansion of the spheroidal harmonic for $s = -2$, is

\begin{eqnarray}
\label{eqc.1}
S_{\ell m} & = & _{-2}Y_{\ell m} + 4a \omega \sum\limits_{\ell \neq \ell'}  \sqrt{\frac{2\ell+1}{2\ell'+1}}
\frac{C^{\ell' m}_{\ell 1 m 0} C^{\ell' m}_{\ell 1 2 0} \  _{-2}Y_{\ell m} }
{[\ell(\ell+1) - \ell'(\ell'+1)]}
\nonumber \\ 
&&
+ O(a^2 \omega^2) \ ,
\end{eqnarray}

\noindent where $C^{a \alpha}_{b \beta c \gamma}$ are the usual Clebsch-Gordon coefficients. For both the horizon mode and the lowest order QNMs, $a \omega < 1$ is true for all $a/M$, and so the expansion is not obviously divergent, although it is only good when $a/M \ll 1$. The inclusion of additional QNM frequencies with $m = 2$ does not remove the residual oscillation in the extraction of $M$ and $a$ seen in Figures \ref{fig:MExtractNocorrect} and \ref{fig:aExtractNocorrect} (though a corotating $\ell = 3, \  m=2$ reduces the amplitude of the oscillation somewhat). In fact, extractions using a large number of modes generally have sharp features, in addition to systematic deviations from the values of $M$ and $a$ given in \cite{Scheel}.

The presence of the $\ell=4,\  m=4$ mode in the $\ell =2,\  m=2$ waveform is unexpected, and we attribute it to errors arising from the numerical generation and extraction of the waveform. The spectral code used in \cite{Scheel} generates its gauge dynamically, and while the waveform extraction method attempts remove gauge effects, studies find that these gauge effects still generate errors \cite{GaugeErrors}. We suspect such gauge errors are the source of mode-mode mixing.

\end{document}